\documentclass[twocolumn,letterpaper,aps,prc,longbibliography,superscriptaddress,showpacs,floatfix]{revtex4-1}

\usepackage{graphicx}	
\usepackage{natbib}
\usepackage{multirow}
\usepackage[toc,page]{appendix}
\usepackage{xspace}	

\newcommand{\pt}{\mbox{$p_T$}\xspace}

\newcommand{\rda}{\mbox{$R_{d{\rm Au}}$}\xspace}

\newcommand{\sqsn}{\mbox{$\sqrt{s_{_{NN}}}$}\xspace}
\newcommand{\pp}{\mbox{$p$$+$$p$}\xspace}
\newcommand{\dau}{\mbox{$d$$+$Au}\xspace}
\newcommand \phee{$\phi \rightarrow e^{+}e^{-}$}
\newcommand \phkk{$\phi \rightarrow K^{+}K^{-}$}
\newcommand \phmm{$\phi \rightarrow \mu^{+}\mu^{-}$}
\newcommand \ee{$e^+e^-$}

\begin{document}


\title{$\phi$ meson production in $d$$+$Au collisions 
       at $\sqrt{s_{_{NN}}}=200$ GeV}

\newcommand{\abilene}{Abilene Christian University, Abilene, Texas 79699, USA}
\newcommand{\augie}{Department of Physics, Augustana College, Sioux Falls, South Dakota 57197, USA}
\newcommand{\banaras}{Department of Physics, Banaras Hindu University, Varanasi 221005, India}
\newcommand{\barc}{Bhabha Atomic Research Centre, Bombay 400 085, India}
\newcommand{\baruch}{Baruch College, City University of New York, New York, New York, 10010 USA}
\newcommand{\bnlcoll}{Collider-Accelerator Department, Brookhaven National Laboratory, Upton, New York 11973\%--5000, USA}
\newcommand{\bnlphys}{Physics Department, Brookhaven National Laboratory, Upton, New York 11973\%--5000, USA}
\newcommand{\caucr}{University of California-Riverside, Riverside, California 92521, USA}
\newcommand{\charlesczech}{Charles University, Ovocn\'{y} trh 5, Praha 1, 116 36, Prague, Czech Republic}
\newcommand{\chonbuk}{Chonbuk National University, Jeonju, 561\%--756, Korea}
\newcommand{\ciae}{Science and Technology on Nuclear Data Laboratory, China Institute of Atomic Energy, Beijing 102413, P.~R.~China}
\newcommand{\cns}{Center for Nuclear Study, Graduate School of Science, University of Tokyo, 7\%--3\%--1 Hongo, Bunkyo, Tokyo 113\%--0033, Japan}
\newcommand{\colorado}{University of Colorado, Boulder, Colorado 80309, USA}
\newcommand{\columbia}{Columbia University, New York, New York 10027 and Nevis Laboratories, Irvington, New York 10533, USA}
\newcommand{\czechtech}{Czech Technical University, Zikova 4, 166 36 Prague 6, Czech Republic}
\newcommand{\dapnia}{Dapnia, CEA Saclay, F-91191, Gif-sur-Yvette, France}
\newcommand{\elte}{ELTE, E{\"o}tv{\"o}s Lor{\'a}nd University, H-1117 Budapest, P{\'a}zm{\'a}ny P.~s.~1/A, Hungary}
\newcommand{\ewha}{Ewha Womans University, Seoul 120-750, Korea}
\newcommand{\fit}{Florida Institute of Technology, Melbourne, Florida 32901, USA}
\newcommand{\fsu}{Florida State University, Tallahassee, Florida 32306, USA}
\newcommand{\gsu}{Georgia State University, Atlanta, Georgia 30303, USA}
\newcommand{\hiroshima}{Hiroshima University, Kagamiyama, Higashi-Hiroshima 739\%--8526, Japan}
\newcommand{\howard}{Department of Physics and Astronomy, Howard University, Washington, DC 20059, USA}
\newcommand{\ihepprot}{IHEP Protvino, State Research Center of Russian Federation, Institute for High Energy Physics, Protvino, 142281, Russia}
\newcommand{\illuiuc}{University of Illinois at Urbana-Champaign, Urbana, Illinois 61801, USA}
\newcommand{\inrras}{Institute for Nuclear Research of the Russian Academy of Sciences, prospekt 60\%--letiya Oktyabrya 7a, Moscow 117312, Russia}
\newcommand{\instpasczech}{Institute of Physics, Academy of Sciences of the Czech Republic, Na Slovance 2, 182 21 Prague 8, Czech Republic}
\newcommand{\isu}{Iowa State University, Ames, Iowa 50011, USA}
\newcommand{\jaea}{Advanced Science Research Center, Japan Atomic Energy Agency, 2\%--4 Shirakata Shirane, Tokai-mura, Naka-gun, Ibaraki-ken 319\%--1195, Japan}
\newcommand{\jyvaskyla}{Helsinki Institute of Physics and University of Jyv{\"a}skyl{\"a}, P.O.Box 35, FI-40014 Jyv{\"a}skyl{\"a}, Finland}
\newcommand{\kek}{KEK, High Energy Accelerator Research Organization, Tsukuba, Ibaraki 305\%--0801, Japan}
\newcommand{\korea}{Korea University, Seoul, 136\%--701, Korea}
\newcommand{\kurchatov}{Russian Research Center ``Kurchatov Institute", Moscow, 123098 Russia}
\newcommand{\kyoto}{Kyoto University, Kyoto 606\%--8502, Japan}
\newcommand{\labllr}{Laboratoire Leprince-Ringuet, Ecole Polytechnique, CNRS-IN2P3, Route de Saclay, F-91128, Palaiseau, France}
\newcommand{\lahorelums}{Physics Department, Lahore University of Management Sciences, Lahore 54792, Pakistan}
\newcommand{\lawllnl}{Lawrence Livermore National Laboratory, Livermore, California 94550, USA}
\newcommand{\losalamos}{Los Alamos National Laboratory, Los Alamos, New Mexico 87545, USA}
\newcommand{\lpc}{LPC, Universit{\'e} Blaise Pascal, CNRS-IN2P3, Clermont-Fd, 63177 Aubiere Cedex, France}
\newcommand{\lund}{Department of Physics, Lund University, Box 118, SE-221 00 Lund, Sweden}
\newcommand{\maryland}{University of Maryland, College Park, Maryland 20742, USA}
\newcommand{\mass}{Department of Physics, University of Massachusetts, Amherst, Massachusetts 01003\%--9337, USA}
\newcommand{\michigan}{Department of Physics, University of Michigan, Ann Arbor, Michigan 48109\%--1040, USA}
\newcommand{\muenster}{Institut f\"ur Kernphysik, University of Muenster, D-48149 Muenster, Germany}
\newcommand{\muhlenberg}{Muhlenberg College, Allentown, Pennsylvania 18104\%--5586, USA}
\newcommand{\myongji}{Myongji University, Yongin, Kyonggido 449\%--728, Korea}
\newcommand{\nagasaki}{Nagasaki Institute of Applied Science, Nagasaki-shi, Nagasaki 851\%--0193, Japan}
\newcommand{\natmephi}{National Research Nuclear University, MEPhI, Moscow Engineering Physics Institute, Moscow, 115409, Russia}
\newcommand{\newmex}{University of New Mexico, Albuquerque, New Mexico 87131, USA}
\newcommand{\nmsu}{New Mexico State University, Las Cruces, New Mexico 88003, USA}
\newcommand{\ohio}{Department of Physics and Astronomy, Ohio University, Athens, Ohio 45701, USA}
\newcommand{\ornl}{Oak Ridge National Laboratory, Oak Ridge, Tennessee 37831, USA}
\newcommand{\orsay}{IPN-Orsay, Universite Paris Sud, CNRS-IN2P3, BP1, F-91406, Orsay, France}
\newcommand{\peking}{Peking University, Beijing 100871, P.~R.~China}
\newcommand{\pnpi}{PNPI, Petersburg Nuclear Physics Institute, Gatchina, Leningrad region, 188300, Russia}
\newcommand{\riken}{RIKEN Nishina Center for Accelerator-Based Science, Wako, Saitama 351\%--0198, Japan}
\newcommand{\rikjrbrc}{RIKEN BNL Research Center, Brookhaven National Laboratory, Upton, New York 11973\%--5000, USA}
\newcommand{\rikkyo}{Physics Department, Rikkyo University, 3\%--34\%--1 Nishi-Ikebukuro, Toshima, Tokyo 171\%--8501, Japan}
\newcommand{\saispbstu}{Saint Petersburg State Polytechnic University, St.~Petersburg, 195251 Russia}
\newcommand{\saopaulo}{Universidade de S{\~a}o Paulo, Instituto de F\'{\i}sica, Caixa Postal 66318, S{\~a}o Paulo CEP05315\%--970, Brazil}
\newcommand{\seoulnat}{Department of Physics and Astronomy, Seoul National University, Seoul 151\%--742, Korea}
\newcommand{\stonybrkc}{Chemistry Department, Stony Brook University, SUNY, Stony Brook, New York 11794\%--3400, USA}
\newcommand{\stonycrkp}{Department of Physics and Astronomy, Stony Brook University, SUNY, Stony Brook, New York 11794\%--3800, USA}
\newcommand{\tenn}{University of Tennessee, Knoxville, Tennessee 37996, USA}
\newcommand{\titech}{Department of Physics, Tokyo Institute of Technology, Oh-okayama, Meguro, Tokyo 152\%--8551, Japan}
\newcommand{\tsukuba}{Institute of Physics, University of Tsukuba, Tsukuba, Ibaraki 305, Japan}
\newcommand{\vandy}{Vanderbilt University, Nashville, Tennessee 37235, USA}
\newcommand{\waseda}{Waseda University, Advanced Research Institute for Science and Engineering, 17  Kikui-cho, Shinjuku-ku, Tokyo 162\%--0044, Japan}
\newcommand{\weizmann}{Weizmann Institute, Rehovot 76100, Israel}
\newcommand{\wigner}{Institute for Particle and Nuclear Physics, Wigner Research Centre for Physics, Hungarian Academy of Sciences (Wigner RCP, RMKI) H-1525 Budapest 114, POBox 49, Budapest, Hungary}
\newcommand{\yonsei}{Yonsei University, IPAP, Seoul 120-749, Korea}
\newcommand{\zagreb}{University of Zagreb, Faculty of Science, Department of Physics, Bijeni\v{c}ka 32, HR-10002 Zagreb, Croatia}
\affiliation{\abilene}
\affiliation{\augie}
\affiliation{\banaras}
\affiliation{\barc}
\affiliation{\baruch}
\affiliation{\bnlcoll}
\affiliation{\bnlphys}
\affiliation{\caucr}
\affiliation{\charlesczech}
\affiliation{\chonbuk}
\affiliation{\ciae}
\affiliation{\cns}
\affiliation{\colorado}
\affiliation{\columbia}
\affiliation{\czechtech}
\affiliation{\dapnia}
\affiliation{\elte}
\affiliation{\ewha}
\affiliation{\fit}
\affiliation{\fsu}
\affiliation{\gsu}
\affiliation{\hiroshima}
\affiliation{\howard}
\affiliation{\ihepprot}
\affiliation{\illuiuc}
\affiliation{\inrras}
\affiliation{\instpasczech}
\affiliation{\isu}
\affiliation{\jaea}
\affiliation{\jyvaskyla}
\affiliation{\kek}
\affiliation{\korea}
\affiliation{\kurchatov}
\affiliation{\kyoto}
\affiliation{\labllr}
\affiliation{\lahorelums}
\affiliation{\lawllnl}
\affiliation{\losalamos}
\affiliation{\lpc}
\affiliation{\lund}
\affiliation{\maryland}
\affiliation{\mass}
\affiliation{\michigan}
\affiliation{\muenster}
\affiliation{\muhlenberg}
\affiliation{\myongji}
\affiliation{\nagasaki}
\affiliation{\natmephi}
\affiliation{\newmex}
\affiliation{\nmsu}
\affiliation{\ohio}
\affiliation{\ornl}
\affiliation{\orsay}
\affiliation{\peking}
\affiliation{\pnpi}
\affiliation{\riken}
\affiliation{\rikjrbrc}
\affiliation{\rikkyo}
\affiliation{\saispbstu}
\affiliation{\saopaulo}
\affiliation{\seoulnat}
\affiliation{\stonybrkc}
\affiliation{\stonycrkp}
\affiliation{\tenn}
\affiliation{\titech}
\affiliation{\tsukuba}
\affiliation{\vandy}
\affiliation{\waseda}
\affiliation{\weizmann}
\affiliation{\wigner}
\affiliation{\yonsei}
\affiliation{\zagreb}
\author{A.~Adare} \affiliation{\colorado} 
\author{C.~Aidala} \affiliation{\mass} \affiliation{\michigan} 
\author{N.N.~Ajitanand} \affiliation{\stonybrkc} 
\author{Y.~Akiba} \affiliation{\riken} \affiliation{\rikjrbrc} 
\author{H.~Al-Bataineh} \affiliation{\nmsu} 
\author{J.~Alexander} \affiliation{\stonybrkc} 
\author{M.~Alfred} \affiliation{\howard} 
\author{A.~Angerami} \affiliation{\columbia} 
\author{K.~Aoki} \affiliation{\kek} \affiliation{\kyoto} \affiliation{\riken} 
\author{N.~Apadula} \affiliation{\isu} \affiliation{\stonycrkp} 
% ...
\collaboration{PHENIX Collaboration} \noaffiliation

\date{\today}



\begin{abstract}

The PHENIX collaboration has measured $\phi$ meson production in 
$d$$+$Au collisions at $\sqrt{s_{_{NN}}}=200$~GeV using the dimuon and 
dielectron decay channels. The $\phi$ meson is measured in the forward 
(backward) $d$-going (Au-going) direction, $1.2<y<2.2$ ($-2.2<y<-1.2$) 
in the transverse-momentum ($p_T$) range from 1--7~GeV/$c$, and at 
midrapidity $|y|<0.35$ in the $p_T$ range below 7~GeV/$c$. The $\phi$ 
meson invariant yields and nuclear-modification factors as 
a function of $p_T$, rapidity, and centrality are reported. An 
enhancement of $\phi$ meson production is observed in the Au-going 
direction, while suppression is seen in the $d$-going direction, and 
no modification is observed at midrapidity relative to the yield in 
$p$$+$$p$ collisions scaled by the number of binary collisions. 
Similar behavior was previously observed for inclusive charged hadrons 
and open heavy flavor indicating similar cold-nuclear-matter effects.


\end{abstract}

	
\maketitle

\section{Introduction}

Collisions of deuterons with gold nuclei (\dau) are of significant
interest in the study of the strongly coupled Quark Gluon Plasma (QGP)
produced at the Relativistic Heavy Ion Collider
(RHIC)~\cite{Adcox:2004mh,Adams:2005dq}. The highly
asymmetric collisions of a small projectile and a large target nucleus
provide a way to investigate the initial state of a nucleus-nucleus
collision experimentally, potentially disentangling effects due to QGP formation
from the cold-nuclear-matter effects.  The latter include
modification of the parton distribution functions (PDFs) in the
nucleus relative to those in the
nucleon~\cite{Heckman:2007wk}, initial-state energy
loss~\cite{Vitev:2003xu}, and the so-called Cronin effect. The
Cronin effect refers to the enhancement of high-\pt particle
production in $p$$+A$ collisions relative to that in \pp collisions
scaled by the number of binary collisions and is often attributed
to multiple scattering of the incoming parton inside the target
nucleus~\cite{Cronin:1974zm,Accardi:2003jh,
  Adare:2013piz,Adare:2014keg}. In addition, results
from recent $p$($d$)$+$A collisions at the Large Hadron Collider and
RHIC suggest that long-range correlations, either present in the
initial state or induced by the evolution of the medium, play an
important role even in these small collision
systems~\cite{Back:2003hx, Back:2004mr, Arsene:2004cn, Abelev:2008ab, CMS:2012qk,Aad:2012gla,
  Adare:2013piz,Chatrchyan:2013nka,Aad:2013fja,ABELEV:2013wsa,
  Aad:2014lta,Abelev:2014mda,Adare:2014keg,Adamczyk:2015xjc}. Detailed
studies of particle production systematics in \dau at RHIC may inform
this
question~\cite{Adare:2012qf,Adare:2012yxa,Adare:2012bv,
  Adare:2013esx,Adare:2013ezl,Adare:2013lkk}.

PHENIX has measured the production of identified particles, such as
$\pi$, $K$ and $p$, at midrapidity in \dau and Au$+$Au collisions and
has found intriguing similarities between the $K/\pi$ and $p/\pi$ ratios 
in peripheral Au$+$Au and central \dau
collisions~\cite{Adare:2013esx}. However, it was observed that
the ratio of spectra in peripheral Au$+$Au to those in central \dau
starts above one at low-\pt and trends to a constant value of $\sim
0.65$ for all identified particles at high \pt.  One explanation for
the low-\pt rise is a relative deficit of midrapidity soft particle
yield in \dau collisions compared to Au$+$Au, which could be due to
the participant asymmetry in the \dau collisions producing a rapidity
shift in the peak of particle
production~\cite{Back:2003hx}. Consequently, measuring
identified particles in different regions of rapidity provides a more
complete picture of \dau collisions and sheds light on the
relationship between \pp, \dau and Au$+$Au.  Measuring identified
particle production at forward rapidity also enables access to the
low-$x$ region where nuclear PDFs (nPDFs) are not well known and where
one expects parton-saturation effects to begin to manifest
in modified particle production.

The yield of $\phi$ mesons in high energy heavy ion collisions
provides key information about the QGP, as the yield is potentially
sensitive to medium-induced effects such as strangeness
enhancement~\cite{Koch:1986ud}, a phenomenon associated with soft
particles in bulk matter which can be accessed through the
measurements of $\phi$ meson production~\cite{Shor:1984ui,
  Alessandro:2003gy, Adler:2004hv,
  Afanasiev:2007tv, Abelev:2007rw, Adamova:2006nu, Alt:2008iv,
  Abelev:2008zk, Adare:2010pt, Arnaldi:2011nn}.  PHENIX
measures $\phi$ meson production over a wide rapidity range and for
various collision systems such as \pp, \dau and Au$+$Au. The
production of $\phi$ mesons has already been measured in \pp, \dau,
Cu$+$Cu and Au$+$Au at midrapidity~\cite{Adare:2010fe,
  Adare:2010pt,Adler:2004hv} and in \pp at forward and
backward rapidities~\cite{Adare:2014mgt} over a wide range in
\pt.

In this paper, the production of $\phi$ mesons is determined at forward 
and backward rapidities via dimuons detected in the PHENIX muon 
spectrometers and at midrapidity via dielectrons detected in the PHENIX 
central arms.  Measurements of the $\phi$ meson nuclear-modification 
factor (\rda) in \dau collisions versus rapidity and versus \pt are presented.  
The results presented here are based on \dau collisions at \sqsn=~200~GeV 
recorded in 2008.  The luminosity used in \phmm~analysis corresponds to 
60~nb$^{-1}$ which is equivalent to a nucleon-nucleon integrated 
luminosity of 24~pb$^{-1}$, while the luminosity used in \phee~analysis 
corresponds to 71~nb$^{-1}$ which is equivalent to a nucleon-nucleon 
integrated luminosity of 28~pb$^{-1}$.

\section{Experiment}

\begin{figure}[htb]
\includegraphics[width=1.0\linewidth]{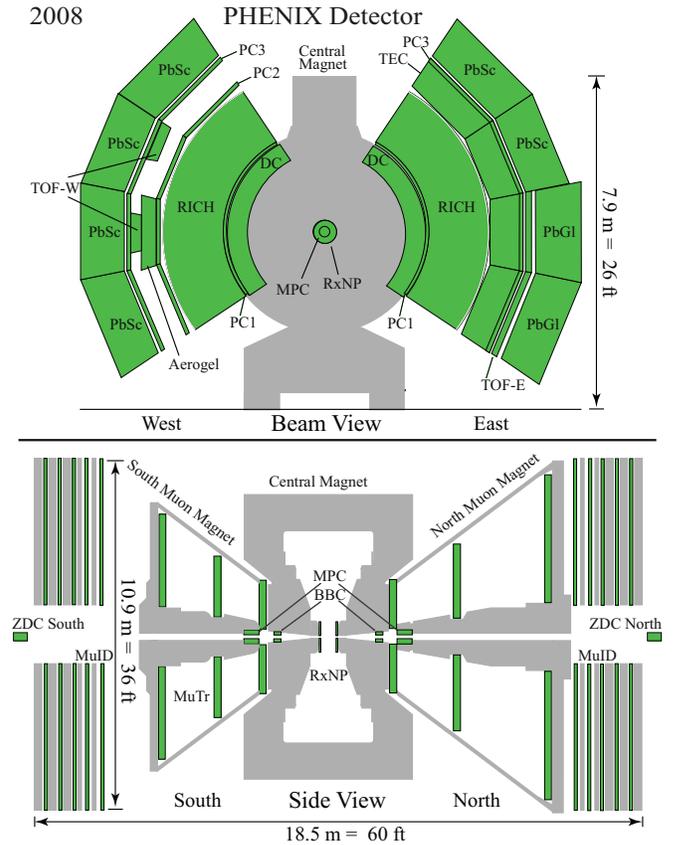}
\caption{\label{fig:det2008} (color online) 
A schematic side view of the PHENIX detector configuration in 
the 2008 data-taking period.}
\end{figure}

Figure~\ref{fig:det2008} shows a schematic of the PHENIX detector, which 
is described in detail in Ref.~\cite{Adcox:2003zm}.  The detectors 
relevant for the analysis of the $\phi$ meson in the dilepton channels are 
the central arms, two muon spectrometers and the two beam-beam counters.

The PHENIX central arms measure particles by using drift chambers (DC)
and pad chambers for the tracking and momentum measurement of charged
particles, ring imaging \v{C}erenkov detectors (RICH) for the
separation of electrons up to the $\pi$ meson \v{C}erenkov radiation
threshold at 5~GeV/$c$, and an electromagnetic calorimeter (EMCal) for
the measurement of spatial positions and energies of photons and
electrons. The EMCal comprises six sectors of lead-scintillator
calorimeter and two sectors of lead-glass calorimeter.

The muon spectrometers, located at forward and backward rapidity, are
shielded by absorbers composed of 19~cm of copper and 60~cm of iron.
Each spectrometer comprises the muon tracker (MuTr) immersed in a
radial magnetic field of integrated bending power of 0.8~T$\cdot$m,
and backed by the muon identifier (MuID). The muon spectrometers
subtend $1.2<|\eta|<2.2$ and the full azimuth. The MuTr comprises
three sets of cathode strip chambers while the MuID comprises five
planes of Iarocci tubes interleaved with steel absorber plates. The
momentum resolution, $\delta p/p$, of particles in the analyzed
momentum range is about 5\%, independent of momentum and dominated by
multiple scattering.  Muon candidates are identified by matching
tracks reconstructed in the MuTr to MuID tracks that penetrate through
to the last MuID plane. The minimum momentum of a muon to reach the
last MuID plane is $\sim$2~GeV/$c$.

Beam-beam counters (BBC), comprising two arrays of 64 \v{C}erenkov
counters covering the pseudorapidity range $3.1<|\eta|<3.9$, measure
the collision vertex along the beam axis ($z_{\rm vtx}$) with 0.5~cm
resolution in addition to providing a minimum-bias (MB) trigger.

\section{Data Analysis}

This section describes the details of the measurement of the $\phi$
meson in \phmm~and \phee~decay channels.

\subsection{\phmm measurement}

The dimuon data set for this analysis was collected in 2008 using a MB
trigger that requires at least one hit in each of the BBC detectors in
conjunction with the MuID Level-1 dimuon trigger.  
Also, at least two tracks were required to penetrate to the last layer of 
the MuID.  The MB trigger measures $88\pm4$\% of the total \dau inelastic 
cross section of 2260~mb~\cite{Adare:2012qf}.

A set of cuts was employed to select good muon candidates and improve
the signal to background ratio. Event selection requires the BBC
collision $z$ vertex to be reconstructed within $\pm$30~cm of the
center of the interaction region along the beam axis. The MuTr tracks
are matched to the MuID tracks at the first MuID layer in both
position and angle. The track is required to have at least 8 of 10
possible hits in the MuID.

The invariant mass distribution is formed by combining muon candidate
tracks of opposite charge. In addition to low mass vector mesons, the
invariant mass spectra in the region of interest contain uncorrelated
and correlated backgrounds.  The uncorrelated backgrounds arise from
random combinatoric associations of unrelated muon candidates while
the correlated backgrounds arise from open charm decay (e.g.,
$D\bar{D}$ where both mesons decay semileptonically to muons), open
bottom decay, $\eta$ and $\omega$ Dalitz decays and from the Drell-Yan
process. The correlated background is much smaller than the
uncorrelated background for all centralities. It is also established
from simulation that the background is dominated by the uncorrelated
contribution from decays, such as $K,\pi \rightarrow \mu\nu$, that occur
in front of and inside the absorber.  A smaller contribution to the
background comes from hadron decays in the muon tracker volume. When
combined into track pairs, these pairs produce a broad distribution of
invariant masses and also a broad distribution of $\chi^{2}_{\rm
  vtx}$, the parameter resulting from fitting the two muon tracks with
the BBC measured event vertex position.  This procedure separates the
foreground and background spectra by applying a cut of $\chi^{2}_{\rm
  vtx}<4.0$ to extract foreground spectra and a cut of $\chi^{2}_{\rm
  vtx}>4.0$ to extract background spectra.  The cut value,
$\chi^{2}_{\rm vtx,cut}=4.0$, is selected such that it retains as much
of the signal as possible within the foreground, while still allowing
sufficient statistics in the background sample. 
This procedure is described in detail in Ref.~\cite{Adare:2014mgt}.

\begin{figure}[htb]
\includegraphics[width=1.0\linewidth]{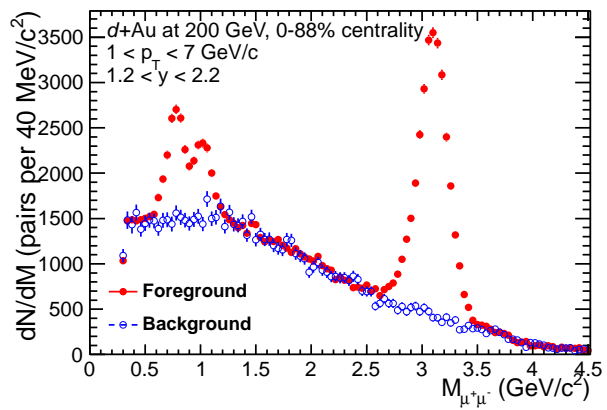}
\caption{\label{fig:bknorm} (color online) 
The unlike-sign dimuon invariant mass spectrum (solid red points) and the 
renormalized background spectrum (open blue circles).}
\end{figure}

To extract the $\phi$ meson signal, the background pairs are
subtracted from the foreground pairs. Because the pairs with
$\chi^{2}_{\rm vtx} > 4.0$ represent only the uncorrelated part of the
background the shape of their mass distribution is slightly altered to
account for the contribution of correlated pairs, which have much
smaller contribution according to simulation results. This is done by
fitting a fourth order polynomial to the ratio of the foreground to
background distributions in the nonresonance region ($0.3 <
M_{\mu^{+}\mu^{-}}<0.5$~GeV/$c^2$ and $1.5 <
M_{\mu^{+}\mu^{-}}<2.5$~GeV/$c^2$). The background distribution is
then multiplied by the polynomial in the entire mass range. 
Figure~~\ref{fig:bknorm} shows the
resulting foreground and renormalized background distributions.

\begin{figure}[htb]
\includegraphics[width=1.0\linewidth]{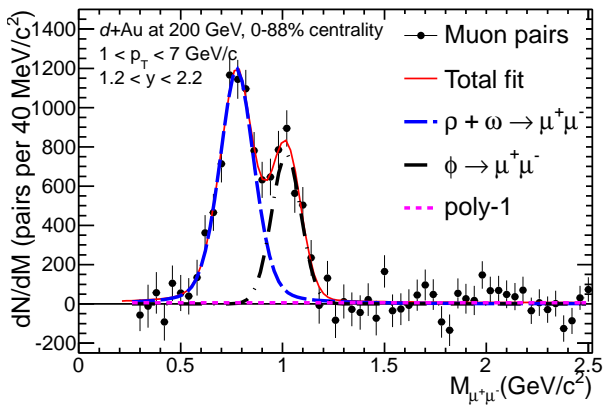}
\caption{\label{fig:northsigbkgdfit} (color online) 
The fitted unlike-sign dimuon spectra after background subtraction.
}
\end{figure}

In the mass range $0.3<M_{\mu^+\mu^-}<2.5$~GeV/$c^2$, the resulting 
distribution after subtraction has contributions from three mesons: 
$\omega$, $\rho$, and $\phi$ mesons. The distribution is fitted with the 
sum of two Gaussian functions and the Breit-Wigner function convoluted 
with a Gaussian. The Gaussian functions are used to fit the $\omega$ and 
$\phi$ mesons while the Breit-Wigner convoluted with a Gaussian is used to 
fit the $\rho$ meson.  Figure~\ref{fig:northsigbkgdfit} shows the results.

The parameters of the fit are fixed to the world average values of the 
masses and widths of the three mesons~\cite{Beringer:1900zz} and the 
widths of the Gaussian distributions, which account for the detector mass 
resolution, are constrained by the values obtained from the simulation. 
The $\phi$ meson mass resolution is $\sim$85~MeV/$c^2$. Because the 
invariant mass peak of the $\phi$ meson is partly resolved in the plot, 
while $\omega$ and $\rho$ meson peaks cannot be resolved, only an estimate 
of the combined yield of these two mesons is allowed. In the fit, the 
ratio of the $\omega$ and $\rho$ mesons, $N_{\rho}/N_\omega$, is set to 
0.58, derived as the ratio of their corresponding production cross 
sections, $\sigma_\rho/\sigma_\omega = 1.15\pm0.15$, 
consistent with values found in jet fragmentation~\cite{JPhysG33:pdg1}, 
multiplied by the ratio of their branching ratios~\cite{Beringer:1900zz}.

The yield extraction is performed in bins of \pt over the range 
$1<p_T<7$~GeV/$c$ for the rapidities $1.2<|y|<2.2$, in 
bins of y for the \pt range $1<p_T<7$~GeV/$c$, and in 
different centrality classes.

The acceptance and reconstruction efficiency ($A\varepsilon_{\rm 
rec}$) of the muon spectrometers, including the MuID trigger 
efficiency, is determined by passing {\sc pythia} 6.421 (default 
parameters)~\cite{Sjostrand:2000wi} generated $\phi$ mesons through 
a full {\sc geant}~\cite{GEANT:W5013} simulation of the PHENIX 
detector.  The {\sc pythia} simulation output is embedded into real \dau data events and then
reconstructed in the same manner as data. Identical cuts to those 
used in the data analysis are applied to this embedded simulation.

\begin{figure}[htb]
\includegraphics[width=1.0\linewidth]{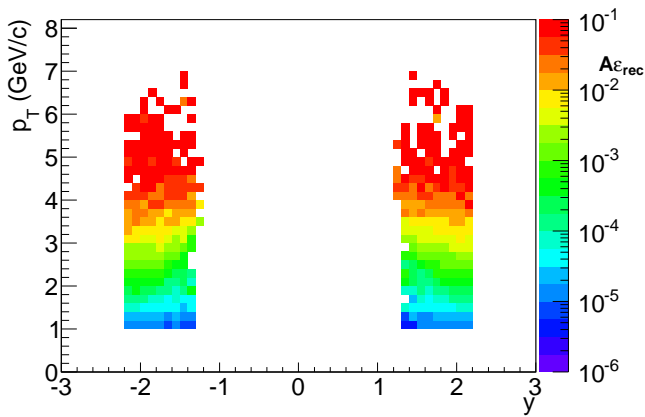}
\caption{\label{fig:AccEff} (color online) The $A\varepsilon_{\rm rec}$ as 
a function of rapidity ($x$-axis) and \pt ($y$-axis) for $\phi$ mesons.}
\end{figure}

Figure~\ref{fig:AccEff} shows the $A\varepsilon_{\rm rec}$ as a function 
of \pt and rapidity for $\phi$ mesons. The $A\varepsilon_{\rm rec}$ 
reaches a few percent at high-\pt and rapidly decreases at low momentum 
limiting this study to $ p_T>1$~GeV/$c$.

\subsection{\phee measurement}

The dielectron data set for this analysis was recorded in 2008 using a MB 
trigger that required at least one hit in each of the BBC detectors with 
an ERT (EMCal-RICH trigger).  The ERT requires a minimum energy deposit of 
0.6 and 0.8~GeV/$c$ in a tile of 2 $\times$ 2 EMCal towers matched to a 
hit in the RICH. The trigger efficiencies are described in detail in 
Ref.~\cite{Adare:2011ht}. 

A set of quality assurance cuts is applied to the data to select good 
electron candidates and improve the signal to background ratio. Event  
selection requires the BBC collision $z$ vertex to be reconstructed 
within $\pm$30~cm of the center of the interaction region along the beam 
direction. Charged tracks are reconstructed using the DC and pad 
chambers and requiring \pt~$>$~0.2~GeV/$c$. Electrons are identified 
mainly by the RICH detector. Furthermore, an electron candidate is 
required to have a good match to a cluster in the EMCAL, and the energy, 
$E$, of the cluster must satisfy the requirement $E/p>0.5$, where $p$ 
is the momentum measured by the DCs. The RICH and EMCal combined provide 
an $e/\pi$ rejection factor of order 10$^4$.

Electrons and positrons reconstructed in an event are combined into pairs. 
The resulting mass spectrum contains both the signal and an inherent 
background. The uncorrelated part of the background is estimated via an 
event-mixing technique~\cite{Adare:2009qk}, which combines electrons from 
different events within the same centrality and vertex class. To extract 
the $\phi$ meson raw yield, the subtracted mass spectrum is fitted by a 
function consisting of several contributions. The $\phi$ and $\omega$ 
meson peaks are approximated by a Breit-Wigner distribution convolved with 
a Gaussian distribution. Parameters of the Breit-Wigner part are set to 
the global averaged values~\cite{Beringer:1900zz} and the width of the 
Gaussian part is to account for the detector mass resolution. Because 
the production ratio of $\rho$ meson to $\omega$ meson is assumed to be 
1.15, their ratio in the fit is given by the ratio of their branching 
ratios to \ee~in vacuum. The $\phi$ meson yield is then extracted by 
summing up the bin contents in a $\pm3\sigma$ window, where $\sigma$ is 
extracted from the fit, and subtracting the background determined by 
integrating the polynomial over the same window.

\begin{figure}[htb]
\includegraphics[width=1.0\linewidth]{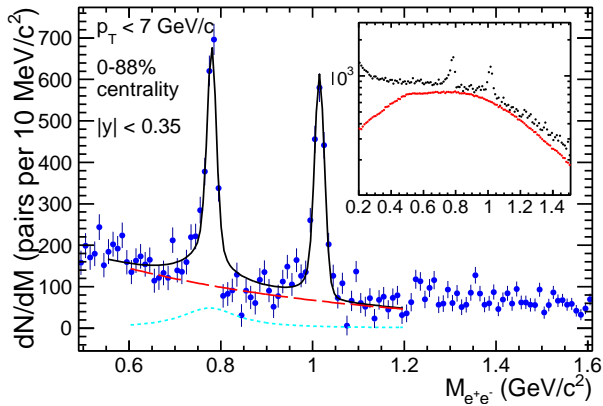}
\caption{\label{fig:phispectra_ee} (color online)
\pt-integrated unlike-sign \ee mass spectrum measured by the PHENIX
central arms after subtracting the uncorrelated combinatorial
background in the minimum-bias \dau collisions. The
dotted-line shows the $\rho$ meson contribution and the dashed line
is the polynomial background, which describes the residual
correlated background.  The insert shows the
raw spectrum before subtraction, overlaid with the normalized
combinatorial background, which is estimated using the mixed-event technique.
}
\end{figure}

Figure~\ref{fig:phispectra_ee} shows an example of the fit to the mixed 
event subtracted mass spectrum. The mixed events were normalized in the 
mass range between 0.7 to 1.5 GeV/$c^2$. The stability of the results 
was checked by varying the normalization region, and the difference was 
included in the systematic errors. The ratio of real to normalized mixed 
events was found to be flat in the region of interest for this analysis 
indicating the validity of the normalization. Complete details describing the estimation 
of the systematic uncertainty on the raw yield extraction by varying the 
background normalization, fitting functions, range and counting method can 
be found in Ref.~\cite{Adare:2011ht}.

The detector acceptance and reconstruction efficiency, as well as 
trigger efficiency were determined using a 
{\sc geant}~\cite{GEANT:W5013} based simulation of the PHENIX detector 
that tuned the detector response to a set of characteristics (dead 
and hot channel maps, gains, noise, etc) that described the 
performance of each detector subsystem. The \phee~ decays were 
generated with a realistic \pt distribution of $\phi$ mesons, within 
$|y|<0.5$ in rapidity, and in full azimuth. The simulated $\phi$ 
meson mean and width of the invariant mass peak were verified to 
match the measured values in real data at all \pt. More details are 
given in Ref.~\cite{Adare:2010fe}.

\subsection{Calculation of invariant yields and nuclear-modification factors}

The dilepton $\phi$ meson invariant yield in a given rapidity, \pt, and 
centrality bin is
\begin{equation}
\frac{B_{ll}}{2 \pi p_T} \frac{d^2 N}{dydp_T} = \frac{1}{2\pi p_T}\frac{1}{\Delta y\Delta p_T}
                          \frac{c N_\phi}{\varepsilon_{\rm tot}N_{\rm evt}},
\end{equation}
where $B_{ll}$ is the $\phi \rightarrow l^{+} l^{-}$ branching ratio, $N_\phi$ is the
measured $\phi$ meson yield, $N_{\rm evt}$ is the number of sampled MB events in the given 
centrality bin, $\Delta y$ is the width of the rapidity bin, $\Delta p_T$ is the width 
of the \pt bin, , and $c$ is the BBC bias correction factor~\cite{Adare:2012qf,
Adare:2013nff}. $\varepsilon_{\rm tot}$ includes trigger efficiencies, acceptance 
and reconstruction efficiency~\cite{Adler:2005ph,Adare:2011ht}. 
The data points are corrected to account for the finite width of the analyzed 
\pt bins~\cite{Lafferty:1994cj}. 

 To gain insight into nuclear medium effects and particle production mechanisms 
in \dau collisions, the ratio of the $\phi$ meson yields in \dau 
collisions to \pp collisions scaled by the number of nucleon-nucleon collisions 
in the \dau system, $N_{\rm coll}$~\cite{Adare:2012qf}, is calculated as: 
\begin{equation}\label{eqn:rdau}
  R_{d{\rm Au}} = \frac{d^2N_{d{\rm Au}}/dydp_T}{N_{\rm coll}\times d^2N_{pp}/dydp_T},
\end{equation}

where $d^2 N_{d{\rm Au}}/dydp_{T}$ is the per-event yield of particle 
production in \dau collisions and $d^2 N_{pp}/dydp_{T}$ is the per-event 
yield of the same process in \pp collisions. The $p$$+$$p$ invariant yield 
used in the \rda calculation for the $\mu^{+}\mu^{-}$~decay channel is the 
$p$$+$$p$ differential cross section from Ref.~\cite{Adare:2014mgt} 
divided by the $p$$+$$p$ total cross section, 42.2~mb. The measured 
$p$$+$$p$ differential cross sections for $e^+e^-$~and $K^+K^-$~decay 
channels are consistent~\cite{Adare:2010fe} and the $p$$+$$p$ reference 
used in the $e^+e^-$~\rda calculations is extracted from a Tsallis 
function fit of the combined $e^+e^-$~and $K^+K^-$~spectra.

\subsection{Systematic uncertainties}

The systematic uncertainties associated with the measured invariant yields 
and nuclear modification factors can be divided into three categories 
based upon the effect each source has on the measured results. All 
uncertainties are reported as standard deviations. Type-A: point-to-point 
uncorrelated uncertainties that allow the data points to move 
independently with respect to one another and are added in quadrature with 
statistical uncertainties. Type-B: point-to-point correlated uncertainties 
that allow the data points to move coherently within the quoted range. 
Type-C: correlated uncertainties that allow the data points to move 
together by a common multiplicative factor, a global uncertainty.

At forward and backward rapidities, a 5\% type-A uncertainty is assigned 
to signal extraction, which corresponds to the average variation between 
the results from the different yield extraction fits. Type-B 
uncertainties, point-to-point correlated uncertainties to some degree in 
\pt, include a 4\% uncertainty from MuID tube efficiency and 2\% from MuTr 
overall efficiency.  An 8\% uncertainty on the yield is assigned to 
account for a 2\% absolute momentum scale uncertainty, which was estimated 
by measuring the $J/\psi$ mass.

\begin{table}[htb]
\caption{
Systematic uncertainties included in the invariant yield and the nuclear 
modification factor calculations at forward and backward rapidities. As 
explained in the text, there is an 8\% type-B systematic uncertainty due 
to small acceptance that impacts the low-\pt region only which is not listed below.
}
\begin{ruledtabular} \begin{tabular}{lcc} 
Source                                 & Value (\%)     & Type \\
\hline
Signal extraction                      & 5              & A \\
MuID efficiency                        & 4              & B \\
MuTr efficiency                        & 2              & B \\
$A\varepsilon_{\rm rec}$ [north/south]  & 8 / 9          & B \\
Absolute momentum scale                & 8              & B \\
Total Type-B                           & 13             & B \\
BBC bias correction for \pp (\dau)     & 10 (0.1--5.8) & C \\
$N_{\rm coll}$                              & 5--7          & C \\
\end{tabular} \end{ruledtabular}
\label{tab:sysUncer}
\caption{
Systematic uncertainties included in the invariant yield and the nuclear 
modification factor calculations at midrapidity.
}
\begin{ruledtabular} \begin{tabular}{lcc} 
Source                                 & Value (\%)    & Type \\
\hline
Signal extraction                      & 8--15         & A \\
ERT trigger efficiency                 & 1--7          & B \\
Acceptance correction                  & 7             & B \\
Electron ID                            & 9             & B \\
Absolute momentum scale                & 1--5          & B \\
Quadratic sum of (B)                   & 11--14        & B \\
BBC bias correction for \pp (\dau)     & 10 (0.1--5.8) & C \\
$N_{\rm coll}$                              & 5--7          & C \\
\end{tabular} \end{ruledtabular}
\label{tab:midsysUncer}
\end{table}

A 9\% (8\%) uncertainty is assigned to the $-2.2<y<-1.2$ ($1.2<y<2.2$) 
rapidity bins due to the uncertainties in the $A\varepsilon_{\rm rec}$ 
determination method itself. The $A\varepsilon_{\rm rec}$ at the lowest \pt bin 
is small and sensitive to variations in the slope of the input \pt distribution 
which affects the differential cross section calculations at this \pt bin. To 
understand this effect, the \pt dependent cross section is fitted by three 
commonly used fit functions (Hagedorn~\cite{Hagedorn:1965st}, 
Kaplan~\cite{Kaplan:1977kr}, and Tsallis~\cite{Adare:2010fe}) 
over the \pt range, $2<p_T<7$~GeV/$c$, and the fitted functions are extrapolated 
to the lowest \pt bin, $1<p_T<2$~GeV/$c$. The differences between the values 
extracted from these fits and the measured one at the lowest \pt bin is within 
8\%, hence an additional 8\% systematic uncertainty is assigned to the lowest 
\pt bin to account for these differences. For the \pt integrated and rapidity 
dependent invariant yields the 8\% uncertainty is assigned to all data bins. 
Type-B systematic uncertainties are added in quadrature and are shown as boxes. 
Finally, an overall normalization uncertainty of 10\% (0.1\%--5.8\%) for \pp 
(\dau) is assigned to the BBC bias correction ($c$) uncertainties and are 
labeled as type-C~\cite{Adare:2012bv}. The BBC bias correction uncertainties 
for \dau vary with centrality and are considered as type-B for centrality dependent spectra. 

 For the nuclear modification factor, the type-A systematic uncertainties 
arise from the quadratic sum of type-A systematic uncertainty in \pp and \dau invariant 
yields~\cite{Adare:2014mgt}. Systematic uncertainties including MuID and 
MuTr efficiencies and absolute momentum scale are the same between the \dau and \pp 
invariant yields and cancel out. A 9\% (7\%) systematic uncertainty 
in the $A\varepsilon_{\rm rec}$ at $-2.2 < y < -1.2$ $(1.2 < y < 2.2)$, that is 
carried over from \pp, is added in quadrature to type-B systematic uncertainties 
listed in Table~\ref{tab:sysUncer}. The acceptance limitation at the lowest \pt bin 
is the same between \dau and \pp data because it is collected using the same 
detector and the associated type-B systematic uncertainty cancels out. 
Type-C systematic uncertainties for the nuclear modification factor are the 
quadratic sum of the type-C systematic uncertainties for the invariant yields 
of \pp and \dau collisions and the uncertainty associated with $N_{\rm coll}$.
The systematic uncertainties are listed in 
Table~\ref{tab:sysUncer}. 

At midrapidity, the main contribution to the systematic uncertainties of 
type-A is the uncertainty associated with the raw yield extraction, 
8\%--15\%. This uncertainty is calculated by varying the background 
normalization, fitting functions, range, and counting methods. Type-B 
systematic uncertainties, point-to-point correlated uncertainties in \pt, 
include uncertainties in the ERT trigger efficiency (1\%--7\%) and 
electron identification (9\%), which are estimated by varying the analysis 
cuts. The acceptance correction uncertainty is 7\% and the momentum scale 
uncertainty is 1\%--5\%.  An overall normalization uncertainty, type-C, 
was assigned for the \pp (\dau) BBC bias correction, which amounts to 10\% 
(0.1\%--5.8\%).

For the nuclear-modification factor, the electron identification and 
branching ratio uncertainties are the same between \pp and \dau and cancel 
out. As for the rest of the systematic uncertainties, the uncertainties 
for each type are added in quadrature between \pp and \dau. Additionally, 
the $N_{\rm coll}$ uncertainty is added in quadrature to the rest of 
type-C uncertainties. The systematic uncertainties are listed in 
Table~\ref{tab:midsysUncer}.

\begin{figure}[ht]
\includegraphics[width=1.0\linewidth]{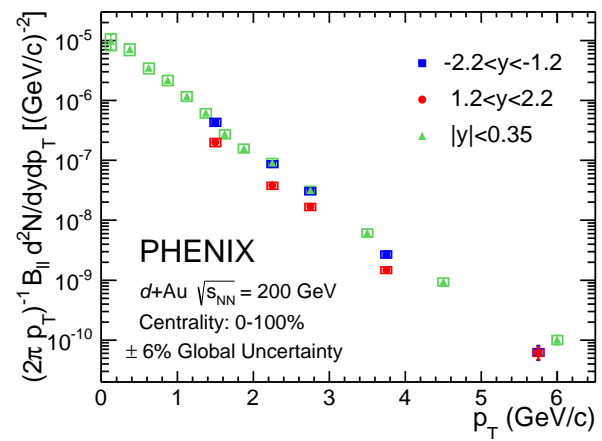}
\caption{\label{fig:dNdydptOmgPhi} (color online) $\phi$ meson
  invariant yields as a function of \pt in the Au-going direction
  (solid blue squares) and in the $d$-going direction (solid red
  points).  The $\phi$ meson invariant yields at midrapidity are shown
  as solid green triangles. The vertical bars represent the
  statistical uncertainties and the boxes represent type-B systematic
  uncertainties.  The ${\pm}6$\% global uncertainty is the associated
  type-C uncertainty.}
\end{figure}

\section{Results}

Figure~\ref{fig:dNdydptOmgPhi} shows the $\phi$ meson invariant yields
in \dau and \pp as a function of \pt in the Au-going direction
$-2.2<y<-1.2$, in the $d$-going direction $1.2<y<2.2$, 
and at midrapidity $|y| < 0.35$. The invariant yields 
measured at forward and backward rapidities for all 
centralities are significantly different from each other. 
At midrapidity, the magnitude of the invariant yield is 
close to 


\begin{figure*}[tbh]
  \begin{minipage}{0.48\linewidth}
\includegraphics[width=0.58\linewidth]{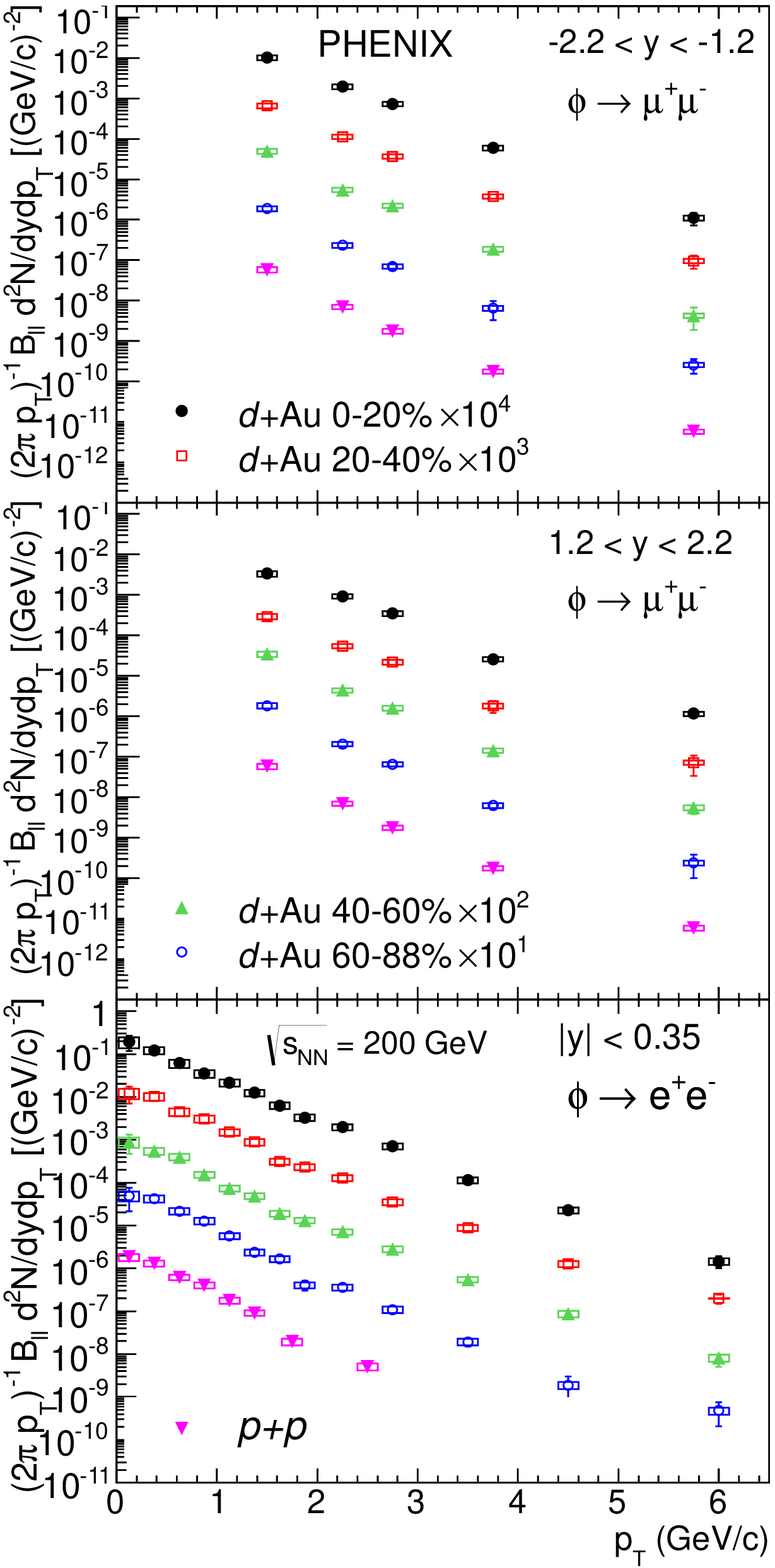}
\caption{\label{fig:ctptdNdydpt} (color online) Invariant \pt spectra
  of the $\phi$ meson for different centrality classes in \dau and
  $p$$+$$p$ collisions at \sqsn~=~200~GeV~\cite{Adare:2014mgt}.
  The vertical bars represent the statistical uncertainties and the
  boxes represent type-B systematic uncertainties. Type-C systematic
  uncertainties are 0.1\%--5.8\% for \dau invariant yields and 10\%
  for \pp invariant yields.  The spectra are scaled by arbitrary
  factors for clarity.  }
  \end{minipage}
  \hspace{0.02\linewidth}
  \begin{minipage}{0.48\linewidth}
\includegraphics[width=0.58\linewidth]{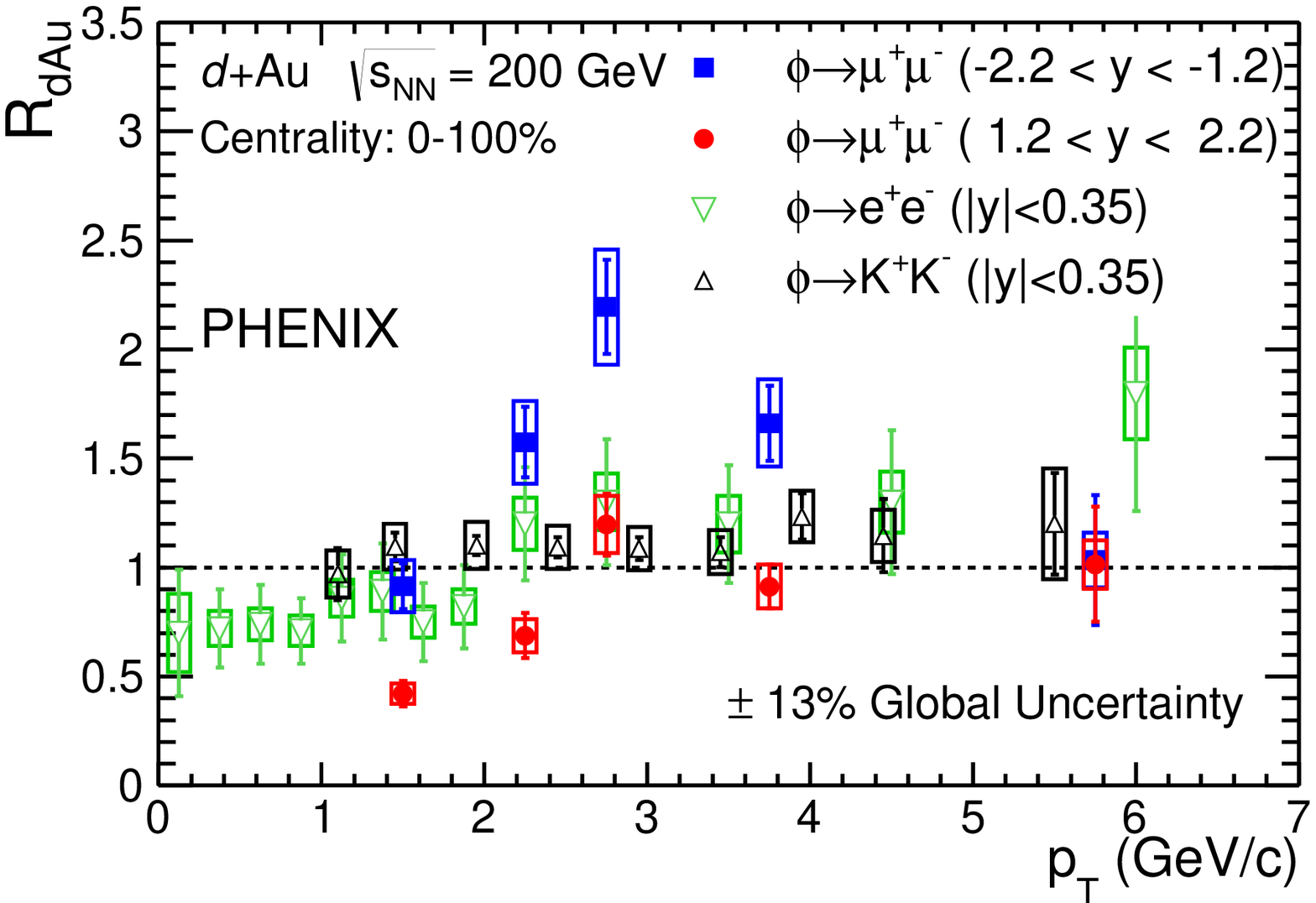}
\caption{\label{fig:opRdAuvspt} (color online) The $\phi$ meson
  nuclear-modification factor, \rda, as a function of \pt.  The solid
  blue squares indicate the Au-going direction and the solid red
  circles indicate the $d$-going direction. The upright black
  triangles are for \phkk at midrapidity~\cite{Adare:2010pt}
  while the inverted triangles are for \phee. The vertical bars
  represent the statistical uncertainties and the boxes represent
  type-B systematic uncertainties. The ${\pm}13$\% global uncertainty
  is the associated type-C uncertainty.  }

\includegraphics[width=0.58\linewidth]{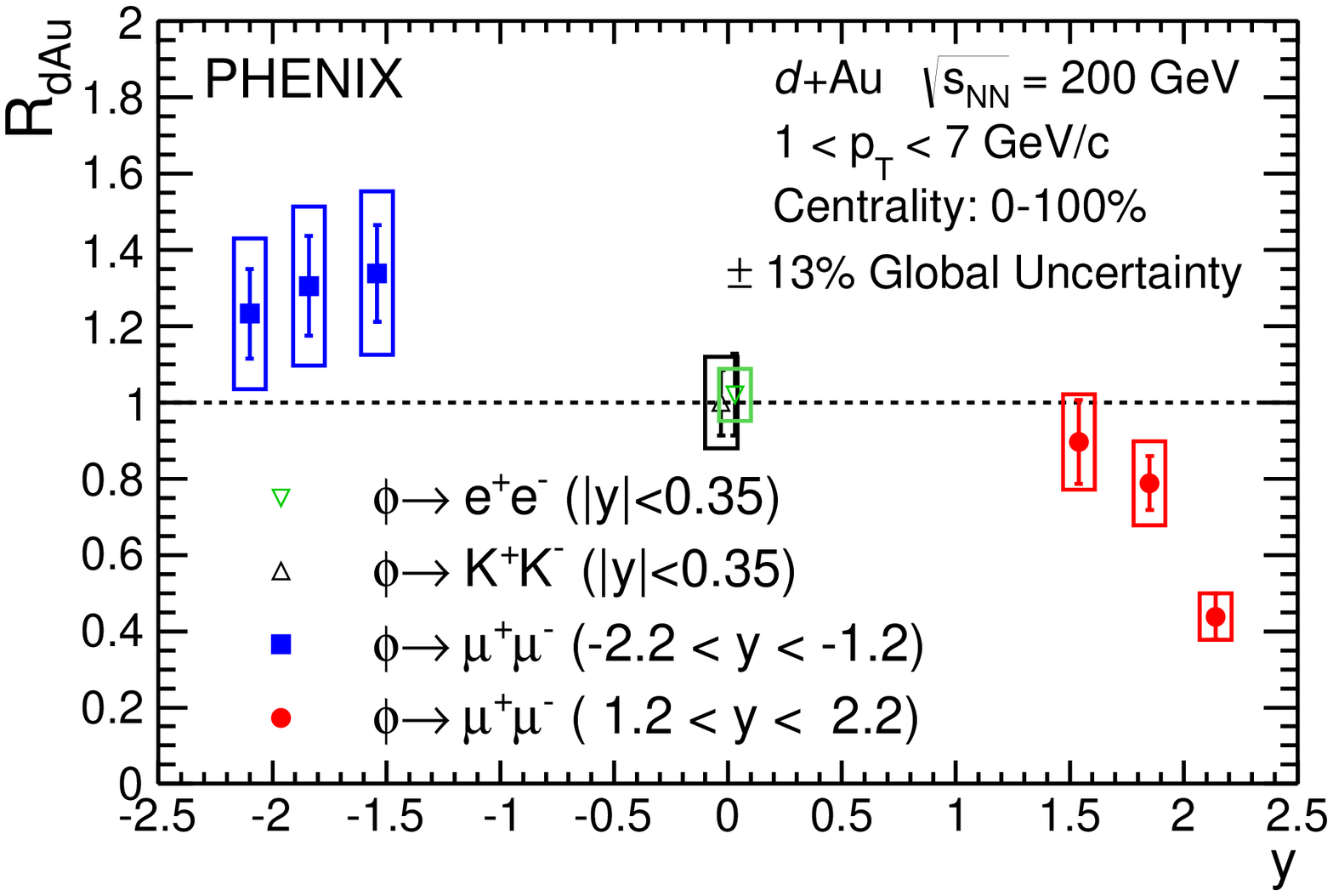}
\caption{\label{fig:opRapRdAu} (color online) The $\phi$ meson nuclear
  modification factor, \rda, as a function of rapidity.  The solid
  blue squares indicate the Au-going direction while the $d$-going
  direction is shown in solid red circles. At midrapidity, the upright
  black triangles are for \phkk~\cite{Adare:2010pt} while the
  inverted triangles are for \phee. The midrapidity points are
  slightly displaced from zero for clarity. The vertical bars
  represent the statistical uncertainties and the boxes represent
  type-B systematic uncertainties. The ${\pm}13$\% global uncertainty
  is the associated type-C uncertainty.  }
  \end{minipage}
\end{figure*}


\noindent that of the backward rapidity, in contrast to what was 
observed in \pp collisions~\cite{Adare:2010fe} where the 
forward and backward rapidities have the same invariant yields and both 
are smaller than the midrapidity invariant yield.

\begin{figure*}[tbh]
  \begin{minipage}{0.48\linewidth}
\includegraphics[width=0.58\linewidth]{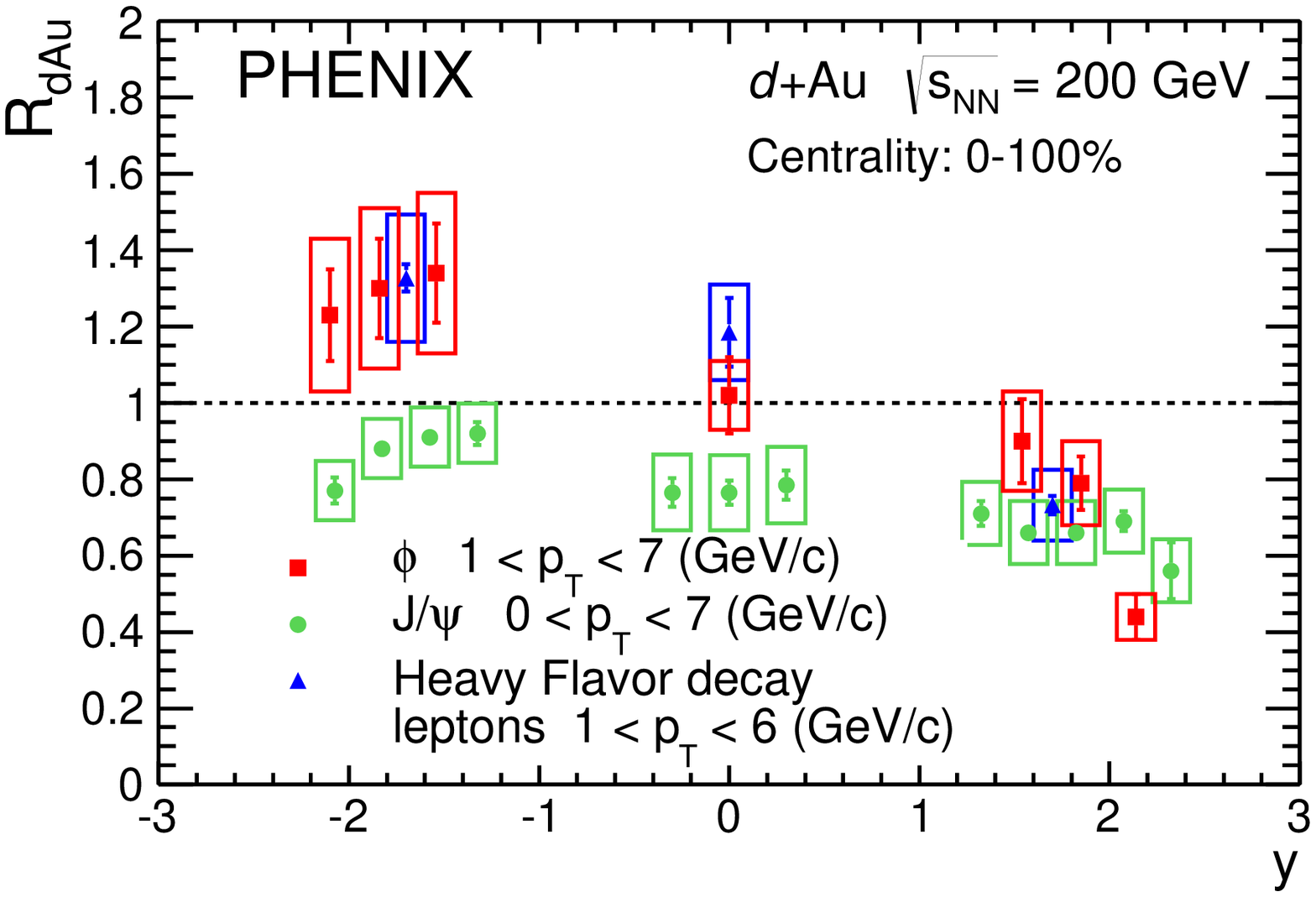}
\caption{\label{fig:jpsihfRapRdAu} (color online)
  $J/\psi$~\cite{Adare:2010fn} (full green circles), heavy
  flavor decay
  leptons~\cite{Adare:2012yxa,Adare:2013lkk} (full
  blue triangles) and $\phi$ meson (full red squares) nuclear
  modification factors, \rda, as a function of rapidity.  The vertical
  bars represent the statistical uncertainties and the boxes represent
  type-B systematic uncertainties. The type-C systematic uncertainties
  associated with heavy-flavor and $J/\psi$ meson measurements are
  10\% and 8\%, respectively.  }
\includegraphics[width=0.58\linewidth]{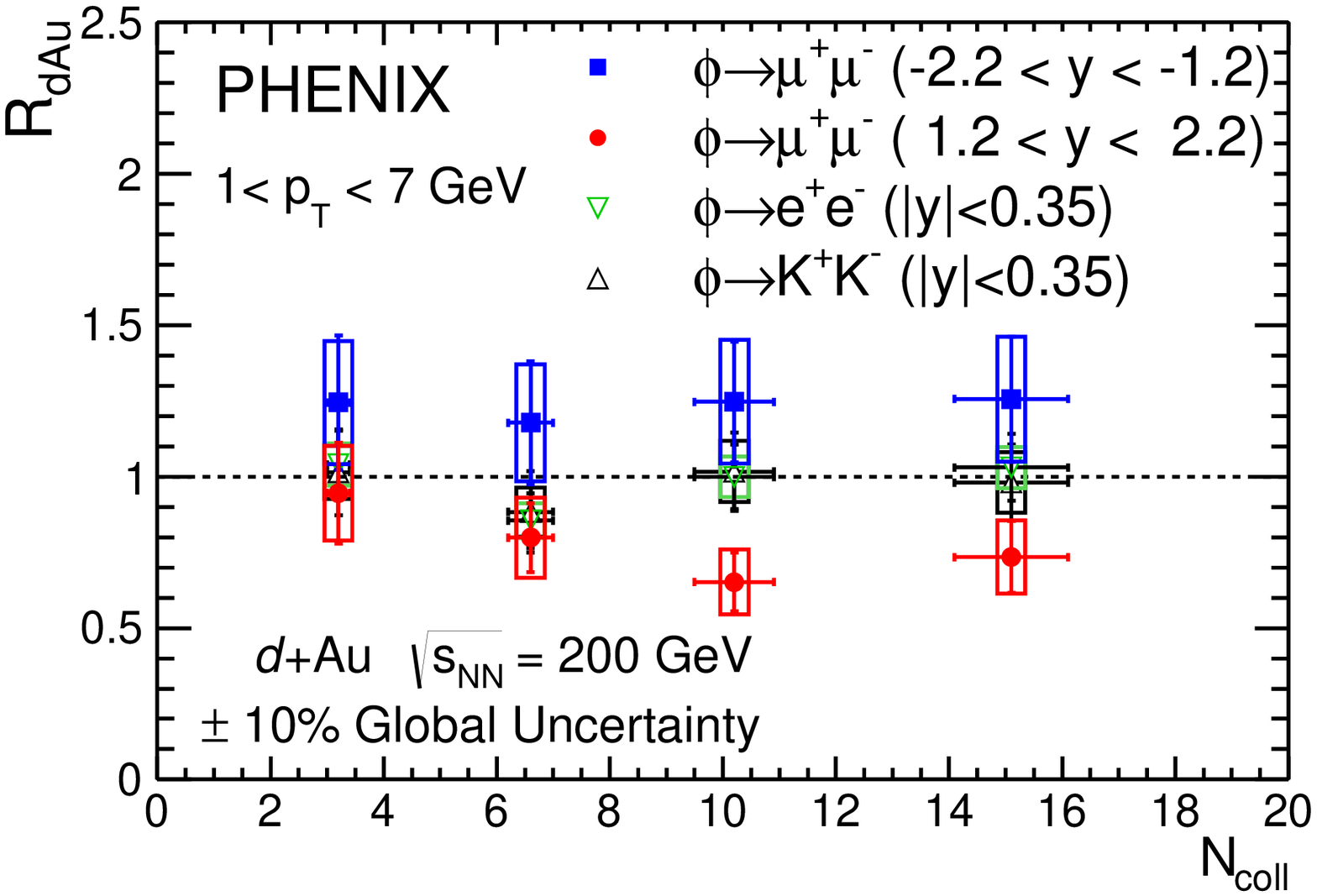}
\caption{\label{fig:opCtRdAu} (color online) The $\phi$ meson nuclear
  modification factor, \rda, as a function of $N_{\rm coll}$.  The
  solid blue squares indicate the Au-going direction while the
  $d$-going direction is shown in solid red circles.  At midrapidity,
  the upright black triangles are for \phkk~\cite{Adare:2010pt}
  while the inverted triangles are for \phee. The vertical bars
  represent the statistical uncertainties and the boxes represent
  type-B systematic uncertainties.  The ${\pm}10$\% global uncertainty
  is the associated type-C uncertainty.  }
  \end{minipage}
  \hspace{0.02\linewidth}
  \begin{minipage}{0.48\linewidth}
\includegraphics[width=0.58\linewidth]{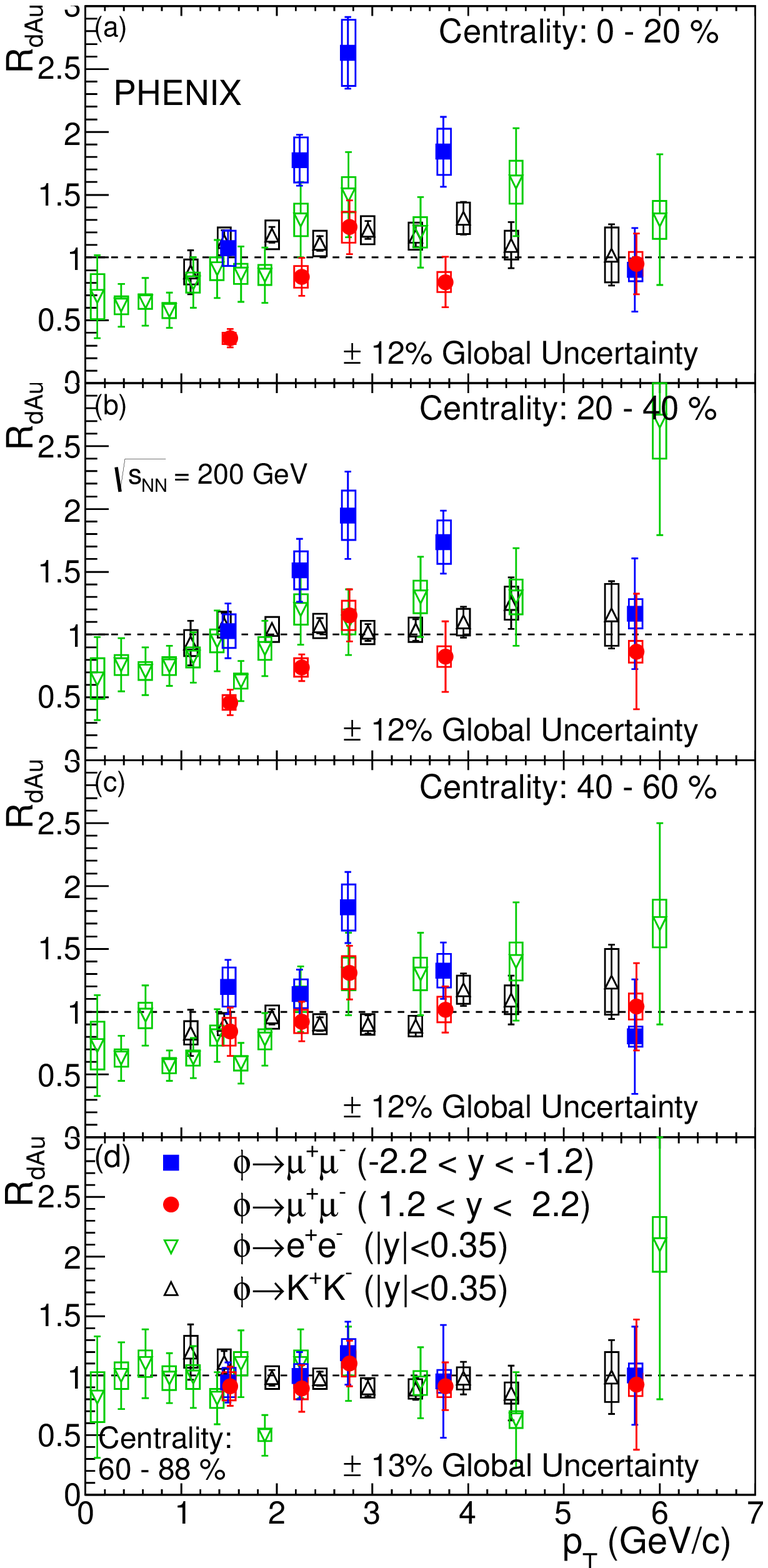}
\caption{\label{fig:ctptRdAuTheo} (color online) The $\phi$ meson \rda
  as a function of \pt for the centralities (a) 0\%--20\%, (b)
  20\%--40\%, (c) 40\%--60\%, and (d) 60\%--88\% in the Au-going
  direction (solid blue squares) and the $d$-going direction (solid
  red circles). At midrapidity, the upright black triangles are from
  \phkk~\cite{Adare:2010pt} while the upside down triangles are
  from \phee. The vertical bars represent the statistical uncertainties
  and the boxes represent type-B systematic uncertainties. The
  ${\pm}$12\%--13\% global uncertainty is the associated type-C
  uncertainty.}
  \end{minipage}
\end{figure*}

Figure~\ref{fig:ctptdNdydpt} shows the invariant yields as a function
of \pt for the different centrality intervals measured in \pp and \dau
collisions. The \dau data are presented in different centrality
classes: 0\%--20\%, 20\%--40\%, 40\%--60\%, and 60\%--88\%, which are 
determined using the BBC over the pseudorapidity range $3.1<|\eta|<3.9$, 
for the $1.2<|y|<2.2$ and $|y|<0.35$ regions. The three 
panels describe the three rapidity regions. The same centrality
definition is used as in Ref.~\cite{Adare:2013nff}.  The wide
rapidity coverage of the full PHENIX detector over an extended \pt
range is demonstrated.

Figure~\ref{fig:opRdAuvspt} shows the \rda for $\phi$ mesons measured
in all centralities as a function of \pt at forward and backward
rapidities and at midrapidity, also compared with \rda measured in the
\phkk~decay channel~\cite{Adare:2010pt}.  The $\phi$ meson \rda
behavior measured in the three different rapidity regions is not the
same. In the Au-going direction, the \rda is consistent with unity at
the lowest measured \pt and increases to $R_{d{\rm Au}} \sim 2$ at \pt
= 2.7~GeV/$c$. It then decreases to unity at the highest measured
\pt. In the $d$-going direction strong suppression is observed at the
lowest measured \pt and then the \rda increases to unity at higher
\pt. At midrapidity, where the measurement starts at \pt=0, the
yield of the $\phi$ meson is suppressed below 1~GeV/$c$ and \rda is
consistent with unity above that. The \rda measured in the \phkk
channel agrees with the measurement of \phee within uncertainties in the \pt
region where the measurements overlap.  The enhanced yield observed in
the Au-going direction at intermediate \pt is characteristic of the
Cronin effect~\cite{Cronin:1974zm}: enhancement of high \pt
particle production in \dau collisions relative to scaled \pp
collisions with a corresponding depletion at low \pt.  The $\phi$ enhancement
in the Au-going direction and the suppression in the $d$-going
direction are consistent with what is observed by ALICE in $p$$+$Pb collisions at 
\sqsn=5.02~TeV in $-4.46<y<-2.96$ and $2.03<y<3.53$~\cite{Adam:2015jca}.  

Figure~\ref{fig:opRapRdAu} shows the \rda of the $\phi$ meson as a
function of rapidity, summed over the \pt range $1 < p_T < 7$~GeV/$c$
and integrated over all centralities.  The \rda is enhanced in the
Au-going direction, shows no modification at midrapidity and is
suppressed in the $d$-going direction.

Figure~\ref{fig:jpsihfRapRdAu} shows that the $\phi$ meson nuclear
modification factor, measured as a function of rapidity, is consistent
with that of leptons coming from decays of heavy flavor
particles~\cite{Adare:2012yxa,Adare:2013lkk}.  This
similarity is interesting given that heavy flavor quark production is 
expected to be dominated by hard processes over
the accessed \pt range.   In contrast, one expects a significant contribution to
$\phi$ meson production from soft processes, particularly at low \pt where the yield is dominant.
This backward/forward enhancement/suppression in \dau collisions is also observed in \dau
charged hadron density results measured by PHOBOS~\cite{Back:2004mr}.
The charged hadron result is often considered as a rapidity shift in the Au-going direction
via soft processes.

Figure~\ref{fig:jpsihfRapRdAu} also shows the $J/\psi$
\rda~\cite{Adare:2010fn}. In the case of the $J/\psi$, the
relative \rda modification as a function of rapidity is similar to
that of the $\phi$ meson and heavy flavor decay leptons. However, the
$J/\psi$ suffers from additional suppression at backward and midrapidity,
which could be due to a larger break up cross section or 
effects in the higher-energy-density backward-rapidity region.

At midrapidity, PHENIX measured the nuclear-modification factors of 
protons and of several mesons from the light ($\pi$) meson up to the 
$\phi$ meson with a mass similar to that of protons.  The results show no 
significant dependence of \rda on the mass of the 
particle~\cite{Adare:2013esx}, whereas the measurements reveal a 
significant dependence of \rda on the number of valence quarks. All mesons 
show no or very small enhancement in comparison to protons in the low to 
medium \pt region.

Figure~\ref{fig:opCtRdAu} shows the \rda integrated over the \pt range 
$1<p_T<7$~GeV/$c$ as a function of centrality for three rapidity 
intervals: $-2.2<y<-1.2$, $1.2<y< 2.2$ and $|y|< 0.35$.  The 
\rda is integrated over the \pt range $1<p_T<7$~GeV/$c$ and the 
rapidity ranges $-2.2<y<-1.2$, $1.2<y<2.2$ and $|y|<0.35$. In the 
Au-going direction, the \rda shows an indication of enhancement with 
$N_{\rm coll}$ and in the $d$-going direction shows a strong suppression 
at high $N_{\rm coll}$. At midrapidity, the \rda for $\phi$ mesons 
reconstructed in $e^+e^-$~is consistent with unity at all centralities.  
This is also consistent with the measurement done in the $K^+K^-$~decay 
channel.

Figure~\ref{fig:ctptRdAuTheo} shows \rda as a function of \pt in different 
$d$$+$Au centrality classes for three rapidity regions covered by the 
PHENIX detector: backward, mid, and forward rapidities. In the most 
peripheral collisions (60\%--88\%), shown in the lowest panel of the 
figure, the nuclear-modification factor measured in all three rapidity 
regions is consistent with unity at all measured \pt. In the more central 
collisions the \rda remains at unity only in the midrapidity region for 
\pt above 1~GeV/$c$. This is consistent between the measurements done in 
the $e^+e^-$~and in the $K^+K^-$~decay channels. In the Au-going 
direction, the \rda reaches a maximum at \pt close to 2.7~GeV/$c$ with an 
amplitude that grows towards more central collisions.  At higher \pt the 
\rda diminishes, approaching unity in all measured centrality classes. In 
the $d$-going direction, all points above 2.7~GeV/$c$ are consistent with 
unity. Below \pt=2.7~GeV/$c$ Fig.~\ref{fig:ctptRdAuTheo} shows \rda$<1$, 
with more suppression in the most central collisions compared to other 
centralities.

\clearpage

The enhancement at backward rapidity is consistent with nuclear \pt 
broadening at moderate \pt and gluon antishadowing, while the suppression 
at forward rapidity may suggest gluon shadowing and/or partonic energy 
loss. The \rda enhancement (suppression) at backward (forward) rapidity in 
the most central collisions decreases gradually from central to peripheral 
collisions where for the most peripheral collisions \rda shows no overall 
modification.  Whether the $\phi$ meson is dominated by soft or hard 
processes remains an open question.

\section{Summary and Conclusions}

PHENIX has measured the production of $\phi$ mesons in \dau collisions
at $\sqsn = 200$~GeV via their decay to dimuons and dielectrons. The
$\phi$ meson is measured in the forward, $d$-going direction, $1.2 < y
< 2.2$ in the \pt range from 1 to 7~GeV/$c$, at midrapidity $|y|<0.35$
in the \pt range below 7~GeV/$c$, and in the backward, the Au-going
direction, $-2.2 < y < -1.2$ in the \pt range from 1 to 7~GeV/$c$.

The measurements reveal that the $\phi$ meson yields in \dau compared to 
binary collision scaled \pp at \pt $>$ 2~GeV/$c$ are suppressed in the 
$d$-going direction and enhanced in the Au-going direction. The yield 
measured at midrapidity is consistent with binary collision scaled \pp.  
No significant modification of $\phi$ meson production is observed in the 
most peripheral \dau collisions. With centrality increasing from 
semi-peripheral events to central the \rda shows more pronounced 
suppression in the $d$-going and more increasing enhancement in Au-going 
direction. In that rapidity region, the \rda has a \pt dependence, which 
is characteristic of a Cronin-type effect. A similar enhancement 
(suppression) was observed by ALICE in $p$$+$Pb collisions at \sqsn = 5.02 
TeV~\cite{Adam:2015jca}. The rapidity dependence of the \rda in 
$\phi$-meson production is similar to the open heavy flavor 
modification~\cite{Adare:2013lkk} which may indicate a general rapidity 
shift compared to the symmetric system and/or similar cold-nuclear matter 
effects are present in both the $\phi$ meson and open heavy-flavor 
production.



\section*{ACKNOWLEDGMENTS}   

We thank the staff of the Collider-Accelerator and Physics
Departments at Brookhaven National Laboratory and the staff of
the other PHENIX participating institutions for their vital
contributions.  We acknowledge support from the 
Office of Nuclear Physics in the
Office of Science of the Department of Energy,
the National Science Foundation, 
Abilene Christian University Research Council, 
Research Foundation of SUNY, and
Dean of the College of Arts and Sciences, Vanderbilt University 
(U.S.A),
Ministry of Education, Culture, Sports, Science, and Technology
and the Japan Society for the Promotion of Science (Japan),
Conselho Nacional de Desenvolvimento Cient\'{\i}fico e
Tecnol{\'o}gico and Funda\c c{\~a}o de Amparo {\`a} Pesquisa do
Estado de S{\~a}o Paulo (Brazil),
Natural Science Foundation of China (P.~R.~China),
Ministry of Science, Education, and Sports (Croatia),
Ministry of Education, Youth and Sports (Czech Republic),
Centre National de la Recherche Scientifique, Commissariat
{\`a} l'{\'E}nergie Atomique, and Institut National de Physique
Nucl{\'e}aire et de Physique des Particules (France),
Bundesministerium f\"ur Bildung und Forschung, Deutscher
Akademischer Austausch Dienst, and Alexander von Humboldt Stiftung (Germany),
National Science Fund, OTKA, K\'aroly R\'obert University College, 
and the Ch. Simonyi Fund (Hungary),
Department of Atomic Energy and Department of Science and Technology (India), 
Israel Science Foundation (Israel), 
Basic Science Research Program through NRF of the Ministry of Education (Korea),
Physics Department, Lahore University of Management Sciences (Pakistan),
Ministry of Education and Science, Russian Academy of Sciences,
Federal Agency of Atomic Energy (Russia),
VR and Wallenberg Foundation (Sweden), 
the U.S. Civilian Research and Development Foundation for the
Independent States of the Former Soviet Union, 
the Hungarian American Enterprise Scholarship Fund,
and the US-Israel Binational Science Foundation.

\section*{APPENDIX}

Tables~\ref{tab:fitPtRes}--\ref{tab:ctptRdAvsPt} show the 
numerical values of the measured invariant yields and \rda that
are plotted in Figs.~\ref{fig:dNdydptOmgPhi}--\ref{fig:ctptRdAuTheo}.
As noted in each caption, the  first uncertainty is statistical and
the second is type-B systematic.

\begingroup \squeezetable

\begin{table*}[ht]
\caption{
Invariant yields of $\phi$ meson production as a function of \pt at different 
\dau centrality classes. The first uncertainty is statistical and 
the second is type-B systematic.
}
\begin{ruledtabular} \begin{tabular}{ccccccc}
 $p_T^{\rm min}$ & $p_T^{\rm max}$ & \multicolumn{5}{c}{$\frac{B_{ll}}{2\pi p_T}
\frac{d^2N}{dydp_T}$ (GeV/$c$)$^{-2}$} \\
\multicolumn{2}{c}{(GeV/$c$)} & 0\%--100\% & 0\%--20\% & 20\%--40\% & 40\%--60\% & 60\%--88\% \\
\hline
 & &  \multicolumn{5}{c}{$-2.2<y<-1.2$} \\
1.0 & 2.0 & (4.35$\pm$0.41$\pm$0.65)$\times10^{-7}$ & (1.01$\pm$0.12$\pm$0.15)$\times10^{-6}$ & (6.56$\pm$1.32$\pm$0.98)$\times10^{-7}$ & (4.94$\pm$0.83$\pm$0.74)$\times10^{-7}$ & (1.89$\pm$0.32$\pm$0.28)$\times10^{-7}$\\
2.0 & 2.5 & (8.70$\pm$0.75$\pm$1.13)$\times10^{-8}$ & (1.95$\pm$0.19$\pm$0.25)$\times10^{-7}$ & (1.12$\pm$0.17$\pm$0.15)$\times10^{-7}$ & (5.47$\pm$0.87$\pm$0.71)$\times10^{-8}$ & (2.32$\pm$0.44$\pm$0.30)$\times10^{-8}$\\
2.5 & 3.0 & (3.09$\pm$0.25$\pm$0.40)$\times10^{-8}$ & (7.35$\pm$0.68$\pm$0.95)$\times10^{-8}$ & (3.68$\pm$0.62$\pm$0.48)$\times10^{-8}$ & (2.23$\pm$0.32$\pm$0.29)$\times10^{-8}$ & (7.05$\pm$1.51$\pm$0.92)$\times10^{-9}$\\
3.0 & 4.5 & (2.70$\pm$0.22$\pm$0.35)$\times10^{-9}$ & (5.95$\pm$0.82$\pm$0.77)$\times10^{-9}$ & (3.79$\pm$0.49$\pm$0.49)$\times10^{-9}$ & (1.87$\pm$0.29$\pm$0.24)$\times10^{-9}$ & (6.52$\pm$3.22$\pm$0.85)$\times10^{-10}$\\
4.5 & 7.0 & (6.35$\pm$1.72$\pm$0.83)$\times10^{-11}$& (1.10$\pm$0.39$\pm$0.14)$\times10^{-10}$& (9.61$\pm$3.51$\pm$1.25)$\times10^{-11}$& (4.28$\pm$2.40$\pm$0.56)$\times10^{-11}$& (2.59$\pm$1.04$\pm$0.34)$\times10^{-11}$\\
\\
 & &  \multicolumn{5}{c}{$1.2<y<2.2$} \\
1.0 & 2.0 & (2.00$\pm$0.25$\pm$0.30)$\times10^{-7}$ & (3.37$\pm$0.65$\pm$0.51)$\times10^{-7}$ & (2.93$\pm$0.61$\pm$0.44)$\times10^{-7}$ & (3.47$\pm$0.76$\pm$0.52)$\times10^{-7}$ & (1.82$\pm$0.31$\pm$0.27)$\times10^{-7}$\\
2.0 & 2.5 & (3.80$\pm$0.54$\pm$0.49)$\times10^{-8}$ & (9.30$\pm$1.59$\pm$1.21)$\times10^{-8}$ & (5.47$\pm$0.73$\pm$0.71)$\times10^{-8}$ & (4.42$\pm$0.71$\pm$0.57)$\times10^{-8}$ & (2.07$\pm$0.44$\pm$0.27)$\times10^{-8}$\\
2.5 & 3.0 & (1.68$\pm$0.18$\pm$0.22)$\times10^{-8}$ & (3.47$\pm$0.56$\pm$0.45)$\times10^{-8}$ & (2.18$\pm$0.37$\pm$0.28)$\times10^{-8}$ & (1.60$\pm$0.25$\pm$0.21)$\times10^{-8}$ & (6.52$\pm$1.07$\pm$0.85)$\times10^{-9}$\\
3.0 & 4.5 & (1.48$\pm$0.14$\pm$0.19)$\times10^{-9}$ & (2.60$\pm$0.63$\pm$0.34)$\times10^{-9}$ & (1.80$\pm$0.60$\pm$0.23)$\times10^{-9}$ & (1.44$\pm$0.24$\pm$0.19)$\times10^{-9}$ & (6.24$\pm$1.33$\pm$0.81)$\times10^{-10}$\\
4.5 & 7.0 & (6.23$\pm$1.51$\pm$0.81)$\times10^{-11}$& (1.16$\pm$0.27$\pm$0.15)$\times10^{-10}$& (7.13$\pm$3.73$\pm$0.93)$\times10^{-11}$& (5.55$\pm$1.78$\pm$0.72)$\times10^{-11}$& (2.39$\pm$1.40$\pm$0.31)$\times10^{-11}$\\
\\
 & & \multicolumn{5}{c}{$|y|<0.35$} \\
0.00 & 0.25 & (9.81$\pm$2.88$\pm$3.01)$\times10^{-6}$ & (1.93$\pm$0.76$\pm$0.55)$\times10^{-5}$ & (1.23$\pm$0.53$\pm$0.33)$\times10^{-5}$ & (8.92$\pm$4.22$\pm$2.42)$\times10^{-6}$ & (4.87$\pm$2.74$\pm$1.26)$\times10^{-6}$  \\
0.25 & 0.50 & (7.18$\pm$0.74$\pm$1.57)$\times10^{-6}$ & (1.23$\pm$0.22$\pm$0.23)$\times10^{-5}$ & (1.03$\pm$0.16$\pm$0.22)$\times10^{-5}$ & (5.46$\pm$1.02$\pm$1.05)$\times10^{-6}$ & (4.28$\pm$0.81$\pm$0.81)$\times10^{-6}$\\
0.50 & 0.75 & (3.49$\pm$0.35$\pm$0.71)$\times10^{-6}$ & (6.11$\pm$1.00$\pm$1.30)$\times10^{-6}$ & (4.52$\pm$0.69$\pm$0.90)$\times10^{-6}$ & (3.96$\pm$0.54$\pm$0.69)$\times10^{-6}$ & (2.18$\pm$0.36$\pm$0.39)$\times10^{-6}$\\
0.75 & 1.00 & (2.19$\pm$0.18$\pm$0.42)$\times10^{-6}$ & (3.54$\pm$0.46$\pm$0.70)$\times10^{-6}$ & (3.13$\pm$0.36$\pm$0.58)$\times10^{-6}$ & (1.54$\pm$0.20$\pm$0.25)$\times10^{-6}$ & (1.28$\pm$0.17$\pm$0.22)$\times10^{-6}$\\
1.00 & 1.25 & (1.18$\pm$0.09$\pm$0.21)$\times10^{-6}$ & (2.16$\pm$0.24$\pm$0.40)$\times10^{-6}$ & (1.51$\pm$0.17$\pm$0.27)$\times10^{-6}$ & (7.44$\pm$1.13$\pm$1.21)$\times10^{-7}$ & (5.70$\pm$0.93$\pm$0.93)$\times10^{-7}$\\
1.25 & 1.50 & (6.20$\pm$0.49$\pm$1.08)$\times10^{-7}$ & (1.26$\pm$0.14$\pm$0.22)$\times10^{-6}$ & (8.92$\pm$0.99$\pm$1.51)$\times10^{-7}$ & (4.87$\pm$0.67$\pm$0.77)$\times10^{-7}$ & (2.37$\pm$0.38$\pm$0.39)$\times10^{-7}$\\
1.50 & 1.75 & (2.75$\pm$0.24$\pm$0.46)$\times10^{-7}$ & (6.32$\pm$0.78$\pm$1.08)$\times10^{-7}$ & (3.13$\pm$0.45$\pm$0.51)$\times10^{-7}$ & (1.90$\pm$0.30$\pm$0.30)$\times10^{-7}$ & (1.65$\pm$0.26$\pm$0.27)$\times10^{-7}$\\
1.75 & 2.00 & (1.59$\pm$0.15$\pm$0.26)$\times10^{-7}$ & (3.31$\pm$0.43$\pm$0.55)$\times10^{-7}$ & (2.34$\pm$0.30$\pm$0.37)$\times10^{-7}$ & (1.32$\pm$0.21$\pm$0.21)$\times10^{-7}$ & (4.11$\pm$1.03$\pm$0.67)$\times10^{-8}$\\
2.00 & 2.50 & (9.22$\pm$0.62$\pm$1.48)$\times10^{-8}$ & (2.01$\pm$0.18$\pm$0.32)$\times10^{-7}$ & (1.29$\pm$0.12$\pm$0.20)$\times10^{-7}$ & (7.15$\pm$0.85$\pm$1.13)$\times10^{-8}$ & (3.60$\pm$0.56$\pm$0.59)$\times10^{-8}$\\
2.50 & 3.00 & (3.19$\pm$0.24$\pm$0.49)$\times10^{-8}$ & (7.12$\pm$0.71$\pm$1.10)$\times10^{-8}$ & (3.54$\pm$0.44$\pm$0.53)$\times10^{-8}$ & (2.84$\pm$0.39$\pm$0.44)$\times10^{-8}$ & (1.10$\pm$0.22$\pm$0.18)$\times10^{-8}$\\
3.00 & 4.00 & (6.17$\pm$0.49$\pm$0.94)$\times10^{-9}$ & (1.15$\pm$0.14$\pm$0.18)$\times10^{-8}$ & (8.92$\pm$1.10$\pm$1.36)$\times10^{-9}$ & (5.52$\pm$0.81$\pm$0.88)$\times10^{-9}$ & (1.94$\pm$0.44$\pm$0.33)$\times10^{-9}$\\
4.00 & 5.00 & (9.45$\pm$1.25$\pm$1.47)$\times10^{-10}$& (2.27$\pm$0.37$\pm$0.37)$\times10^{-9}$ & (1.27$\pm$0.26$\pm$0.20)$\times10^{-9}$ & (8.74$\pm$2.06$\pm$1.57)$\times10^{-10}$& (1.87$\pm$1.10$\pm$0.34)$\times10^{-10}$\\
5.00 & 7.00 & (1.02$\pm$0.19$\pm$0.17)$\times10^{-10}$& (1.46$\pm$0.45$\pm$0.25)$\times10^{-10}$& (1.98$\pm$0.50$\pm$0.03)$\times10^{-10}$& (8.27$\pm$3.16$\pm$1.59)$\times10^{-11}$& (4.79$\pm$2.74$\pm$0.90)$\times10^{-11}$\\
\end{tabular} \end{ruledtabular}
\label{tab:fitPtRes}
\end{table*}

\endgroup

\begin{table*}[ht]
\caption{
\rda as a function of rapidity of $\phi$ meson summed over the \pt range, 
$1<p_T<7$~GeV/$c$ for 0\%--100\% centrality. The first uncertainty is 
statistical and the second is type-B systematic.
}
\begin{ruledtabular} \begin{tabular}{cccccccc} 
& & $y_{\rm min}$ & $y_{\rm max}$ & \rda & & \\
\hline
& &  -2.2 & -2.0  & 1.23$\pm$0.12$\pm$0.20 & & \\
& &  -2.0 & -1.7  & 1.30$\pm$0.13$\pm$0.21 & & \\
& &  -1.7 & -1.1  & 1.34$\pm$0.13$\pm$0.21 & & \\
& & -0.35 & 0.35  & 1.02$\pm$0.11$\pm$0.07 & & \\
& &   1.2 & 1.7   & 0.90$\pm$0.11$\pm$0.13 & & \\
& &   1.7 & 2.0   & 0.79$\pm$0.07$\pm$0.11 & & \\
& &   2.0 & 2.2   & 0.44$\pm$0.06$\pm$0.06 & & \\
\end{tabular} \end{ruledtabular}
\label{tab:RdAuvy}
\end{table*}


\begin{table*}[ht]
\caption{
\rda vs $N_{\rm coll}$ of $\phi$ meson at $1 < p_T < 7$~GeV/$c$. The first 
uncertainty is statistical and the second is type-B systematic.
}
\begin{ruledtabular} \begin{tabular}{ccccccc} 
 & & & \multicolumn{3}{c}{\rda} & \\
& Centrality & $\langle N_{\rm coll}  \rangle$ & $-2.2<y<-1.2$ & $1.2<y<2.2$ & $|y|<0.35$ & \\
\hline
& 0\%--20\% & 15.1$\pm$1.0 & 1.26$\pm$0.21$\pm$0.21 & 0.73$\pm$0.12$\pm$0.12 & 1.03$\pm$0.11$\pm$0.09 & \\
& 20\%--40\% & 10.2$\pm$0.7 & 1.25$\pm$0.20$\pm$0.20 & 0.65$\pm$0.10$\pm$0.11 & 1.00$\pm$0.10$\pm$0.09 & \\
& 40\%--60\% &  6.6$\pm$0.4 & 1.18$\pm$0.20$\pm$0.19 & 0.80$\pm$0.11$\pm$0.13 & 0.86$\pm$0.09$\pm$0.08 & \\
& 60\%--88\% &  3.2$\pm$0.2 & 1.24$\pm$0.22$\pm$0.20 & 0.95$\pm$0.17$\pm$0.16 & 1.04$\pm$0.11$\pm$0.09 & \\
\end{tabular} \end{ruledtabular}
\label{tab:RdAuvct}
\end{table*}

\begin{table*}[ht]
\caption{
\rda and \pt at different \dau centrality classes. The first uncertainty 
is statistical and the second is type-B systematic.
}
\begin{ruledtabular} \begin{tabular}{ccccccc} 
 $p_T^{\rm min}$ & $p_T^{\rm max}$ & \multicolumn{5}{c}{\rda} \\
\\
\multicolumn{2}{c}{(GeV/$c$)} & 0\%--100\% & 0\%--20\% & 20\%--40\% & 40\%--60\% & 60\%--88\% \\
\hline
 & & \multicolumn{5}{c}{$-2.2<y<-1.2$} \\
1.0 & 2.0 & 0.92$\pm$0.10$\pm$0.12 & 1.07$\pm$0.14$\pm$0.14 & 1.03$\pm$0.22$\pm$0.13 & 1.20$\pm$0.22$\pm$0.16 & 0.94$\pm$0.17$\pm$0.12\\
2.0 & 2.5 & 1.57$\pm$0.16$\pm$0.19 & 1.78$\pm$0.20$\pm$0.18 & 1.51$\pm$0.25$\pm$0.15 & 1.14$\pm$0.19$\pm$0.11 & 1.00$\pm$0.20$\pm$0.10\\
2.5 & 3.0 & 2.19$\pm$0.22$\pm$0.26 & 2.63$\pm$0.29$\pm$0.26 & 1.95$\pm$0.35$\pm$0.19 & 1.83$\pm$0.28$\pm$0.18 & 1.19$\pm$0.26$\pm$0.12\\
3.0 & 4.5 & 1.66$\pm$0.17$\pm$0.20 & 1.84$\pm$0.28$\pm$0.18 & 1.74$\pm$0.25$\pm$0.17 & 1.33$\pm$0.22$\pm$0.13 & 0.95$\pm$0.47$\pm$0.10\\
4.5 & 7.0 & 1.03$\pm$0.30$\pm$0.12 & 0.90$\pm$0.33$\pm$0.09 & 1.17$\pm$0.44$\pm$0.12 & 0.80$\pm$0.46$\pm$0.08 & 1.00$\pm$0.41$\pm$0.10\\
\\
 & & \multicolumn{5}{c}{$1.2<y<2.2$} \\
1.0 & 2.0 & 0.42$\pm$0.06$\pm$0.05 & 0.36$\pm$0.07$\pm$0.05 & 0.46$\pm$0.10$\pm$0.06 & 0.84$\pm$0.19$\pm$0.11 & 0.91$\pm$0.16$\pm$0.12\\
2.0 & 2.5 & 0.69$\pm$0.10$\pm$0.08 & 0.85$\pm$0.15$\pm$0.08 & 0.74$\pm$0.11$\pm$0.07 & 0.92$\pm$0.16$\pm$0.09 & 0.89$\pm$0.20$\pm$0.09\\
2.5 & 3.0 & 1.20$\pm$0.14$\pm$0.13 & 1.24$\pm$0.21$\pm$0.12 & 1.15$\pm$0.21$\pm$0.12 & 1.31$\pm$0.22$\pm$0.13 & 1.10$\pm$0.19$\pm$0.11\\
3.0 & 4.5 & 0.91$\pm$0.10$\pm$0.10 & 0.80$\pm$0.20$\pm$0.08 & 0.83$\pm$0.28$\pm$0.08 & 1.02$\pm$0.18$\pm$0.10 & 0.91$\pm$0.20$\pm$0.09\\
4.5 & 7.0 & 1.01$\pm$0.26$\pm$0.11 & 0.95$\pm$0.24$\pm$0.10 & 0.87$\pm$0.46$\pm$0.09 & 1.04$\pm$0.35$\pm$0.10 & 0.92$\pm$0.55$\pm$0.09\\
\\
 & & \multicolumn{5}{c}{$|y|<0.35$} \\
0.00 & 0.25 & 0.70$\pm$0.29$\pm$0.18 & 0.69$\pm$0.33$\pm$0.18 & 0.65$\pm$0.33$\pm$0.16 & 0.73$\pm$0.40$\pm$0.19 & 0.82$\pm$0.51$\pm$0.21\\
0.25 & 0.50 & 0.72$\pm$0.18$\pm$0.08 & 0.62$\pm$0.17$\pm$0.07 & 0.76$\pm$0.21$\pm$0.08 & 0.63$\pm$0.18$\pm$0.07 & 1.00$\pm$0.28$\pm$0.11\\
0.50 & 0.75 & 0.74$\pm$0.18$\pm$0.07 & 0.65$\pm$0.19$\pm$0.06 & 0.71$\pm$0.19$\pm$0.07 & 0.97$\pm$0.24$\pm$0.10 & 1.10$\pm$0.29$\pm$0.11\\
0.75 & 1.00 & 0.71$\pm$0.15$\pm$0.07 & 0.58$\pm$0.14$\pm$0.06 & 0.75$\pm$0.16$\pm$0.07 & 0.57$\pm$0.12$\pm$0.06 & 0.98$\pm$0.21$\pm$0.10\\
1.00 & 1.25 & 0.86$\pm$0.20$\pm$0.09 & 0.80$\pm$0.20$\pm$0.08 & 0.82$\pm$0.20$\pm$0.08 & 0.63$\pm$0.16$\pm$0.06 & 0.99$\pm$0.26$\pm$0.10\\
1.25 & 1.50 & 0.89$\pm$0.22$\pm$0.09 & 0.91$\pm$0.23$\pm$0.09 & 0.95$\pm$0.24$\pm$0.09 & 0.81$\pm$0.21$\pm$0.08 & 0.81$\pm$0.22$\pm$0.08\\
1.50 & 1.75 & 0.75$\pm$0.18$\pm$0.07 & 0.87$\pm$0.22$\pm$0.08 & 0.63$\pm$0.16$\pm$0.06 & 0.60$\pm$0.16$\pm$0.06 & 1.10$\pm$0.28$\pm$0.10\\
1.75 & 2.00 & 0.82$\pm$0.19$\pm$0.08 & 0.86$\pm$0.22$\pm$0.08 & 0.89$\pm$0.22$\pm$0.09 & 0.78$\pm$0.21$\pm$0.08 & 0.50$\pm$0.17$\pm$0.05\\
2.00 & 2.50 & 1.20$\pm$0.26$\pm$0.12 & 1.30$\pm$0.30$\pm$0.13 & 1.20$\pm$0.28$\pm$0.12 & 1.10$\pm$0.26$\pm$0.10 & 1.10$\pm$0.29$\pm$0.11\\
2.50 & 3.00 & 1.30$\pm$0.29$\pm$0.13 & 1.50$\pm$0.34$\pm$0.14 & 1.10$\pm$0.26$\pm$0.10 & 1.30$\pm$0.33$\pm$0.13 & 1.10$\pm$0.31$\pm$0.10\\
3.00 & 4.00 & 1.20$\pm$0.27$\pm$0.13 & 1.20$\pm$0.28$\pm$0.12 & 1.30$\pm$0.32$\pm$0.13 & 1.30$\pm$0.33$\pm$0.13 & 0.94$\pm$0.30$\pm$0.09\\
4.00 & 5.00 & 1.30$\pm$0.33$\pm$0.14 & 1.60$\pm$0.43$\pm$0.17 & 1.30$\pm$0.39$\pm$0.14 & 1.40$\pm$0.47$\pm$0.15 & 0.63$\pm$0.40$\pm$0.07\\
5.00 & 7.00 & 1.80$\pm$0.54$\pm$0.21 & 1.30$\pm$0.52$\pm$0.15 & 2.70$\pm$0.91$\pm$0.30 & 1.70$\pm$0.80$\pm$0.19 & 2.10$\pm$1.30$\pm$0.23\\
\end{tabular} \end{ruledtabular}
\label{tab:ctptRdAvsPt}
\end{table*}


\clearpage


\begin{thebibliography}{53}%
\makeatletter
\providecommand \@ifxundefined [1]{%
 \@ifx{#1\undefined}
}%
\providecommand \@ifnum [1]{%
 \ifnum #1\expandafter \@firstoftwo
 \else \expandafter \@secondoftwo
 \fi
}%
\providecommand \@ifx [1]{%
 \ifx #1\expandafter \@firstoftwo
 \else \expandafter \@secondoftwo
 \fi
}%
\providecommand \natexlab [1]{#1}%
\providecommand \enquote  [1]{``#1''}%
\providecommand \bibnamefont  [1]{#1}%
\providecommand \bibfnamefont [1]{#1}%
\providecommand \citenamefont [1]{#1}%
\providecommand \href@noop [0]{\@secondoftwo}%
\providecommand \href [0]{\begingroup \@sanitize@url \@href}%
\providecommand \@href[1]{\@@startlink{#1}\@@href}%
\providecommand \@@href[1]{\endgroup#1\@@endlink}%
\providecommand \@sanitize@url [0]{\catcode `\\12\catcode `\$12\catcode
  `\&12\catcode `\#12\catcode `\^12\catcode `\_12\catcode `\%12\relax}%
\providecommand \@@startlink[1]{}%
\providecommand \@@endlink[0]{}%
\providecommand \url  [0]{\begingroup\@sanitize@url \@url }%
\providecommand \@url [1]{\endgroup\@href {#1}{\urlprefix }}%
\providecommand \urlprefix  [0]{URL }%
\providecommand \Eprint [0]{\href }%
\providecommand \doibase [0]{http://dx.doi.org/}%
\providecommand \selectlanguage [0]{\@gobble}%
\providecommand \bibinfo  [0]{\@secondoftwo}%
\providecommand \bibfield  [0]{\@secondoftwo}%
\providecommand \translation [1]{[#1]}%
\providecommand \BibitemOpen [0]{}%
\providecommand \bibitemStop [0]{}%
\providecommand \bibitemNoStop [0]{.\EOS\space}%
\providecommand \EOS [0]{\spacefactor3000\relax}%
\providecommand \BibitemShut  [1]{\csname bibitem#1\endcsname}%
\let\auto@bib@innerbib\@empty
\bibitem [{\citenamefont {Adcox}\ \emph {et~al.}(2005)\citenamefont {Adcox}
  \emph {et~al.}}]{Adcox:2004mh}%
  \BibitemOpen
  \bibfield  {author} {\bibinfo {author} {\bibfnamefont {K.}~\bibnamefont
  {Adcox}} \emph {et~al.} (\bibinfo {collaboration} {PHENIX Collaboration}),\
  }\bibfield  {title} {\enquote {\bibinfo {title} {{Formation of dense partonic
  matter in relativistic nucleus-nucleus collisions at RHIC: Experimental
  evaluation by the PHENIX Collaboration}},}\ }\href {\doibase
  10.1016/j.nuclphysa.2005.03.086} {\bibfield  {journal} {\bibinfo  {journal}
  {Nucl. Phys. A}\ }\textbf {\bibinfo {volume} {757}},\ \bibinfo {pages} {184}
  (\bibinfo {year} {2005})}\BibitemShut {NoStop}%
\bibitem [{\citenamefont {Adams}\ \emph {et~al.}(2005)\citenamefont {Adams}
  \emph {et~al.}}]{Adams:2005dq}%
  \BibitemOpen
  \bibfield  {author} {\bibinfo {author} {\bibfnamefont {J.}~\bibnamefont
  {Adams}} \emph {et~al.} (\bibinfo {collaboration} {STAR Collaboration}),\
  }\bibfield  {title} {\enquote {\bibinfo {title} {{Experimental and
  theoretical challenges in the search for the quark gluon plasma: The STAR
  Collaboration's critical assessment of the evidence from RHIC collisions}},}\
  }\href {\doibase 10.1016/j.nuclphysa.2005.03.085} {\bibfield  {journal}
  {\bibinfo  {journal} {Nucl. Phys. A}\ }\textbf {\bibinfo {volume} {757}},\
  \bibinfo {pages} {102} (\bibinfo {year} {2005})}\BibitemShut {NoStop}%
\bibitem [{\citenamefont {Heckman}\ \emph {et~al.}(2007)\citenamefont
  {Heckman}, \citenamefont {Seo},\ and\ \citenamefont {Vafa}}]{Heckman:2007wk}%
  \BibitemOpen
  \bibfield  {author} {\bibinfo {author} {\bibfnamefont {J.~J.}\ \bibnamefont
  {Heckman}}, \bibinfo {author} {\bibfnamefont {J.}~\bibnamefont {Seo}}, \ and\
  \bibinfo {author} {\bibfnamefont {C.}~\bibnamefont {Vafa}},\ }\bibfield
  {title} {\enquote {\bibinfo {title} {{Phase Structure of a Brane/Anti-Brane
  System at Large N}},}\ }\href {\doibase 10.1088/1126-6708/2007/07/073}
  {\bibfield  {journal} {\bibinfo  {journal} {J. High Energy Phys.}\ }\textbf
  {\bibinfo {volume} {0707}},\ \bibinfo {pages} {073} (\bibinfo {year}
  {2007})}\BibitemShut {NoStop}%
\bibitem [{\citenamefont {Vitev}(2003)}]{Vitev:2003xu}%
  \BibitemOpen
  \bibfield  {author} {\bibinfo {author} {\bibfnamefont {I.}~\bibnamefont
  {Vitev}},\ }\bibfield  {title} {\enquote {\bibinfo {title} {{Initial state
  parton broadening and energy loss probed in $d$+Au at RHIC}},}\ }\href
  {\doibase 10.1016/S0370-2693(03)00535-5} {\bibfield  {journal} {\bibinfo
  {journal} {Phys. Lett. B}\ }\textbf {\bibinfo {volume} {562}},\ \bibinfo
  {pages} {36} (\bibinfo {year} {2003})}\BibitemShut {NoStop}%
\bibitem [{\citenamefont {Cronin}\ \emph {et~al.}(1975)\citenamefont {Cronin},
  \citenamefont {Frisch}, \citenamefont {Shochet}, \citenamefont {Boymond},
  \citenamefont {Mermod} \emph {et~al.}}]{Cronin:1974zm}%
  \BibitemOpen
  \bibfield  {author} {\bibinfo {author} {\bibfnamefont {J.~W.}\ \bibnamefont
  {Cronin}}, \bibinfo {author} {\bibfnamefont {H.~J.}\ \bibnamefont {Frisch}},
  \bibinfo {author} {\bibfnamefont {M.J.}\ \bibnamefont {Shochet}}, \bibinfo
  {author} {\bibfnamefont {J.P.}\ \bibnamefont {Boymond}}, \bibinfo {author}
  {\bibfnamefont {R.}~\bibnamefont {Mermod}},  \emph {et~al.},\ }\bibfield
  {title} {\enquote {\bibinfo {title} {{Production of Hadrons with Large
  Transverse Momentum at 200~GeV, 300~GeV, and 400~GeV}},}\ }\href {\doibase
  10.1103/PhysRevD.11.3105} {\bibfield  {journal} {\bibinfo  {journal} {Phys.
  Rev. D}\ }\textbf {\bibinfo {volume} {11}},\ \bibinfo {pages} {3105}
  (\bibinfo {year} {1975})}\BibitemShut {NoStop}%
\bibitem [{\citenamefont {Accardi}\ and\ \citenamefont
  {Gyulassy}(2004)}]{Accardi:2003jh}%
  \BibitemOpen
  \bibfield  {author} {\bibinfo {author} {\bibfnamefont {A.}~\bibnamefont
  {Accardi}}\ and\ \bibinfo {author} {\bibfnamefont {M.}~\bibnamefont
  {Gyulassy}},\ }\bibfield  {title} {\enquote {\bibinfo {title} {{Cronin effect
  versus geometrical shadowing in $d$+Au collisions at RHIC}},}\ }\href
  {\doibase 10.1016/j.physletb.2004.02.020} {\bibfield  {journal} {\bibinfo
  {journal} {Phys. Lett. B}\ }\textbf {\bibinfo {volume} {586}},\ \bibinfo
  {pages} {244} (\bibinfo {year} {2004})}\BibitemShut {NoStop}%
\bibitem [{\citenamefont {Adare}\ \emph
  {et~al.}(2013{\natexlab{a}})\citenamefont {Adare} \emph
  {et~al.}}]{Adare:2013piz}%
  \BibitemOpen
  \bibfield  {author} {\bibinfo {author} {\bibfnamefont {A.}~\bibnamefont
  {Adare}} \emph {et~al.} (\bibinfo {collaboration} {PHENIX Collaboration}),\
  }\bibfield  {title} {\enquote {\bibinfo {title} {{Quadrupole Anisotropy in
  Dihadron Azimuthal Correlations in Central $d$$+$Au Collisions at
  $\sqrt{s_{_{NN}}}$=200 GeV}},}\ }\href {\doibase
  10.1103/PhysRevLett.111.212301} {\bibfield  {journal} {\bibinfo  {journal}
  {Phys. Rev. Lett.}\ }\textbf {\bibinfo {volume} {111}},\ \bibinfo {pages}
  {212301} (\bibinfo {year} {2013}{\natexlab{a}})}\BibitemShut {NoStop}%
\bibitem [{\citenamefont {Adare}\ \emph {et~al.}(2015)\citenamefont {Adare}
  \emph {et~al.}}]{Adare:2014keg}%
  \BibitemOpen
  \bibfield  {author} {\bibinfo {author} {\bibfnamefont {A.}~\bibnamefont
  {Adare}} \emph {et~al.} (\bibinfo {collaboration} {PHENIX Collaboration}),\
  }\bibfield  {title} {\enquote {\bibinfo {title} {{Measurement of long-range
  angular correlation and quadrupole anisotropy of pions and (anti)protons in
  central $d$$+$Au collisions at $\sqrt{s_{_{NN}}}$=200 GeV}},}\ }\href
  {\doibase 10.1103/PhysRevLett.114.192301} {\bibfield  {journal} {\bibinfo
  {journal} {Phys. Rev. Lett.}\ }\textbf {\bibinfo {volume} {114}},\ \bibinfo
  {pages} {192301} (\bibinfo {year} {2015})}\BibitemShut {NoStop}%
\bibitem [{\citenamefont {Back}\ \emph {et~al.}(2004)\citenamefont {Back} \emph
  {et~al.}}]{Back:2003hx}%
  \BibitemOpen
  \bibfield  {author} {\bibinfo {author} {\bibfnamefont {B.~B.}\ \bibnamefont
  {Back}} \emph {et~al.} (\bibinfo {collaboration} {PHOBOS Collaboration}),\
  }\bibfield  {title} {\enquote {\bibinfo {title} {{Pseudorapidity distribution
  of charged particles in $d$+Au collisions at $\sqrt{s_{NN}}$=200~GeV}},}\
  }\href {\doibase 10.1103/PhysRevLett.93.082301} {\bibfield  {journal}
  {\bibinfo  {journal} {Phys. Rev. Lett.}\ }\textbf {\bibinfo {volume} {93}},\
  \bibinfo {pages} {082301} (\bibinfo {year} {2004})}\BibitemShut {NoStop}%
\bibitem [{\citenamefont {Back}\ \emph {et~al.}(2005)\citenamefont {Back} \emph
  {et~al.}}]{Back:2004mr}%
  \BibitemOpen
  \bibfield  {author} {\bibinfo {author} {\bibfnamefont {B.~B.}\ \bibnamefont
  {Back}} \emph {et~al.} (\bibinfo {collaboration} {PHOBOS Collaboration}),\
  }\bibfield  {title} {\enquote {\bibinfo {title} {{Scaling of charged particle
  production in $d$+Au collisions at $\sqrt{s_{NN}}$=200~GeV}},}\ }\href
  {\doibase 10.1103/PhysRevC.72.031901} {\bibfield  {journal} {\bibinfo
  {journal} {Phys. Rev. C}\ }\textbf {\bibinfo {volume} {72}},\ \bibinfo
  {pages} {031901} (\bibinfo {year} {2005})}\BibitemShut {NoStop}%
\bibitem [{\citenamefont {Arsene}\ \emph {et~al.}(2005)\citenamefont {Arsene}
  \emph {et~al.}}]{Arsene:2004cn}%
  \BibitemOpen
  \bibfield  {author} {\bibinfo {author} {\bibfnamefont {I.}~\bibnamefont
  {Arsene}} \emph {et~al.} (\bibinfo {collaboration} {BRAHMS Collaboration}),\
  }\bibfield  {title} {\enquote {\bibinfo {title} {{Centrality Dependence of
  Charged-Particle Pseudorapidity Distributions from $d$$+$Au collisions at
  $\sqrt{s_{_{NN}}}$=200 GeV}},}\ }\href {\doibase 10.1103/Arsene:2004cn}
  {\bibfield  {journal} {\bibinfo  {journal} {Phys. Rev. Lett.}\ }\textbf
  {\bibinfo {volume} {94}},\ \bibinfo {pages} {032301} (\bibinfo {year}
  {2005})}\BibitemShut {NoStop}%
\bibitem [{\citenamefont {Abelev}\ \emph
  {et~al.}(2009{\natexlab{a}})\citenamefont {Abelev} \emph
  {et~al.}}]{Abelev:2008ab}%
  \BibitemOpen
  \bibfield  {author} {\bibinfo {author} {\bibfnamefont {B.~I.}\ \bibnamefont
  {Abelev}} \emph {et~al.} (\bibinfo {collaboration} {STAR Collaboration}),\
  }\bibfield  {title} {\enquote {\bibinfo {title} {{Systematic measurements of
  identified particle spectra in pp, d+Au, and Au+Au collisions at the STAR
  detector}},}\ }\href {\doibase 10.1103/Abelev:2008ab} {\bibfield  {journal}
  {\bibinfo  {journal} {Phys. Rev. C}\ }\textbf {\bibinfo {volume} {79}},\
  \bibinfo {pages} {034909} (\bibinfo {year} {2009}{\natexlab{a}})}\BibitemShut
  {NoStop}%
\bibitem [{\citenamefont {Chatrchyan}\ \emph
  {et~al.}(2013{\natexlab{a}})\citenamefont {Chatrchyan} \emph
  {et~al.}}]{CMS:2012qk}%
  \BibitemOpen
  \bibfield  {author} {\bibinfo {author} {\bibfnamefont {S.}~\bibnamefont
  {Chatrchyan}} \emph {et~al.} (\bibinfo {collaboration} {CMS Collaboration}),\
  }\bibfield  {title} {\enquote {\bibinfo {title} {{Observation of long-range
  near-side angular correlations in proton-lead collisions at the LHC}},}\
  }\href {\doibase 10.1016/j.physletb.2012.11.025} {\bibfield  {journal}
  {\bibinfo  {journal} {Phys. Lett. B}\ }\textbf {\bibinfo {volume} {718}},\
  \bibinfo {pages} {795} (\bibinfo {year} {2013}{\natexlab{a}})}\BibitemShut
  {NoStop}%
\bibitem [{\citenamefont {Aad}\ \emph {et~al.}(2013{\natexlab{a}})\citenamefont
  {Aad} \emph {et~al.}}]{Aad:2012gla}%
  \BibitemOpen
  \bibfield  {author} {\bibinfo {author} {\bibfnamefont {G.}~\bibnamefont
  {Aad}} \emph {et~al.} (\bibinfo {collaboration} {ATLAS Collaboration}),\
  }\bibfield  {title} {\enquote {\bibinfo {title} {{Observation of Associated
  Near-Side and Away-Side Long-Range Correlations in $\sqrt{s_{NN}}$=5.02~TeV
  Proton-Lead Collisions with the ATLAS Detector}},}\ }\href {\doibase
  10.1103/PhysRevLett.110.182302} {\bibfield  {journal} {\bibinfo  {journal}
  {Phys. Rev. Lett.}\ }\textbf {\bibinfo {volume} {110}},\ \bibinfo {pages}
  {182302} (\bibinfo {year} {2013}{\natexlab{a}})}\BibitemShut {NoStop}%
\bibitem [{\citenamefont {Chatrchyan}\ \emph
  {et~al.}(2013{\natexlab{b}})\citenamefont {Chatrchyan} \emph
  {et~al.}}]{Chatrchyan:2013nka}%
  \BibitemOpen
  \bibfield  {author} {\bibinfo {author} {\bibfnamefont {S.}~\bibnamefont
  {Chatrchyan}} \emph {et~al.} (\bibinfo {collaboration} {CMS Collaboration}),\
  }\bibfield  {title} {\enquote {\bibinfo {title} {{Multiplicity and transverse
  momentum dependence of two- and four-particle correlations in $p$Pb and PbPb
  collisions}},}\ }\href {\doibase 10.1016/j.physletb.2013.06.028} {\bibfield
  {journal} {\bibinfo  {journal} {Phys. Lett. B}\ }\textbf {\bibinfo {volume}
  {724}},\ \bibinfo {pages} {213} (\bibinfo {year}
  {2013}{\natexlab{b}})}\BibitemShut {NoStop}%
\bibitem [{\citenamefont {Aad}\ \emph {et~al.}(2013{\natexlab{b}})\citenamefont
  {Aad} \emph {et~al.}}]{Aad:2013fja}%
  \BibitemOpen
  \bibfield  {author} {\bibinfo {author} {\bibfnamefont {G.}~\bibnamefont
  {Aad}} \emph {et~al.} (\bibinfo {collaboration} {ATLAS Collaboration}),\
  }\bibfield  {title} {\enquote {\bibinfo {title} {{Measurement with the ATLAS
  detector of multi-particle azimuthal correlations in $p$+Pb collisions at
  $\sqrt{s_{NN}}$=5.02 TeV}},}\ }\href {\doibase
  10.1016/j.physletb.2013.06.057} {\bibfield  {journal} {\bibinfo  {journal}
  {Phys. Lett. B}\ }\textbf {\bibinfo {volume} {725}},\ \bibinfo {pages} {60}
  (\bibinfo {year} {2013}{\natexlab{b}})}\BibitemShut {NoStop}%
\bibitem [{\citenamefont {Abelev}\ \emph {et~al.}(2013)\citenamefont {Abelev}
  \emph {et~al.}}]{ABELEV:2013wsa}%
  \BibitemOpen
  \bibfield  {author} {\bibinfo {author} {\bibfnamefont {B.~B.}\ \bibnamefont
  {Abelev}} \emph {et~al.} (\bibinfo {collaboration} {ALICE Collaboration}),\
  }\bibfield  {title} {\enquote {\bibinfo {title} {{Long-range angular
  correlations of $\pi$, $K$, and $p$ in $p$-Pb collisions at $\sqrt{s_{NN}}$ =
  5.02 TeV}},}\ }\href {\doibase 10.1016/j.physletb.2013.08.024} {\bibfield
  {journal} {\bibinfo  {journal} {Phys. Lett. B}\ }\textbf {\bibinfo {volume}
  {726}},\ \bibinfo {pages} {164} (\bibinfo {year} {2013})}\BibitemShut
  {NoStop}%
\bibitem [{\citenamefont {Aad}\ \emph {et~al.}(2014)\citenamefont {Aad} \emph
  {et~al.}}]{Aad:2014lta}%
  \BibitemOpen
  \bibfield  {author} {\bibinfo {author} {\bibfnamefont {G.}~\bibnamefont
  {Aad}} \emph {et~al.} (\bibinfo {collaboration} {ATLAS Collaboration}),\
  }\bibfield  {title} {\enquote {\bibinfo {title} {{Measurement of long-range
  pseudorapidity correlations and azimuthal harmonics in $\sqrt{s_{NN}}=5.02$
  TeV proton-lead collisions with the ATLAS detector}},}\ }\href {\doibase
  10.1103/PhysRevC.90.044906} {\bibfield  {journal} {\bibinfo  {journal} {Phys.
  Rev. C}\ }\textbf {\bibinfo {volume} {90}},\ \bibinfo {pages} {044906}
  (\bibinfo {year} {2014})}\BibitemShut {NoStop}%
\bibitem [{\citenamefont {Abelev}\ \emph {et~al.}(2014)\citenamefont {Abelev}
  \emph {et~al.}}]{Abelev:2014mda}%
  \BibitemOpen
  \bibfield  {author} {\bibinfo {author} {\bibfnamefont {B.~B.}\ \bibnamefont
  {Abelev}} \emph {et~al.} (\bibinfo {collaboration} {ALICE Collaboration}),\
  }\bibfield  {title} {\enquote {\bibinfo {title} {{Multiparticle azimuthal
  correlations in $p$-Pb and Pb-Pb collisions at the CERN Large Hadron
  Collider}},}\ }\href {\doibase 10.1103/PhysRevC.90.054901} {\bibfield
  {journal} {\bibinfo  {journal} {Phys. Rev. C}\ }\textbf {\bibinfo {volume}
  {90}},\ \bibinfo {pages} {054901} (\bibinfo {year} {2014})}\BibitemShut
  {NoStop}%
\bibitem [{\citenamefont {Adamczyk}\ \emph {et~al.}(2015)\citenamefont
  {Adamczyk} \emph {et~al.}}]{Adamczyk:2015xjc}%
  \BibitemOpen
  \bibfield  {author} {\bibinfo {author} {\bibfnamefont {L.}~\bibnamefont
  {Adamczyk}} \emph {et~al.} (\bibinfo {collaboration} {STAR Collaboration}),\
  }\bibfield  {title} {\enquote {\bibinfo {title} {{Long-range pseudorapidity
  dihadron correlations in $d$$+$Au collisions at $\sqrt{s_{NN}}$=5.02 GeV}},}\
  }\href {\doibase 10.1016/j.physletb.2015.05.075} {\bibfield  {journal}
  {\bibinfo  {journal} {Phys. Lett. B}\ }\textbf {\bibinfo {volume} {747}},\
  \bibinfo {pages} {265} (\bibinfo {year} {2015})}\BibitemShut {NoStop}%
\bibitem [{\citenamefont {Adare}\ \emph
  {et~al.}(2013{\natexlab{b}})\citenamefont {Adare} \emph
  {et~al.}}]{Adare:2012qf}%
  \BibitemOpen
  \bibfield  {author} {\bibinfo {author} {\bibfnamefont {A.}~\bibnamefont
  {Adare}} \emph {et~al.} (\bibinfo {collaboration} {PHENIX Collaboration}),\
  }\bibfield  {title} {\enquote {\bibinfo {title} {{Transverse-Momentum
  Dependence of the $J/\psi$ Nuclear Modification in $d+$Au Collisions at
  $\sqrt{s_{NN}}$=200 GeV}},}\ }\href {\doibase 10.1103/PhysRevC.87.034904}
  {\bibfield  {journal} {\bibinfo  {journal} {Phys. Rev. C}\ }\textbf {\bibinfo
  {volume} {87}},\ \bibinfo {pages} {034904} (\bibinfo {year}
  {2013}{\natexlab{b}})}\BibitemShut {NoStop}%
\bibitem [{\citenamefont {Adare}\ \emph {et~al.}(2012)\citenamefont {Adare}
  \emph {et~al.}}]{Adare:2012yxa}%
  \BibitemOpen
  \bibfield  {author} {\bibinfo {author} {\bibfnamefont {A.}~\bibnamefont
  {Adare}} \emph {et~al.} (\bibinfo {collaboration} {PHENIX Collaboration}),\
  }\bibfield  {title} {\enquote {\bibinfo {title} {{Cold-nuclear-matter effects
  on heavy-quark production in $d+$Au collisions at $\sqrt{s_{NN}}$=200
  GeV}},}\ }\href {\doibase 10.1103/PhysRevLett.109.242301} {\bibfield
  {journal} {\bibinfo  {journal} {Phys. Rev. Lett.}\ }\textbf {\bibinfo
  {volume} {109}},\ \bibinfo {pages} {242301} (\bibinfo {year}
  {2012})}\BibitemShut {NoStop}%
\bibitem [{\citenamefont {Adare}\ \emph
  {et~al.}(2013{\natexlab{c}})\citenamefont {Adare} \emph
  {et~al.}}]{Adare:2012bv}%
  \BibitemOpen
  \bibfield  {author} {\bibinfo {author} {\bibfnamefont {A.}~\bibnamefont
  {Adare}} \emph {et~al.} (\bibinfo {collaboration} {PHENIX Collaboration}),\
  }\bibfield  {title} {\enquote {\bibinfo {title} {{$\Upsilon(1S+2S+3S)$
  production in $d$$+$Au and $p$$+$$p$ collisions at $\sqrt{s_{NN}}$=200 GeV
  and cold-nuclear matter effects}},}\ }\href {\doibase
  10.1103/PhysRevC.87.044909} {\bibfield  {journal} {\bibinfo  {journal} {Phys.
  Rev. C}\ }\textbf {\bibinfo {volume} {87}},\ \bibinfo {pages} {044909}
  (\bibinfo {year} {2013}{\natexlab{c}})}\BibitemShut {NoStop}%
\bibitem [{\citenamefont {Adare}\ \emph
  {et~al.}(2013{\natexlab{d}})\citenamefont {Adare} \emph
  {et~al.}}]{Adare:2013esx}%
  \BibitemOpen
  \bibfield  {author} {\bibinfo {author} {\bibfnamefont {A.}~\bibnamefont
  {Adare}} \emph {et~al.} (\bibinfo {collaboration} {PHENIX Collaboration}),\
  }\bibfield  {title} {\enquote {\bibinfo {title} {{Spectra and ratios of
  identified particles in Au+Au and $d$+Au collisions at $\sqrt{s_{NN}}$=200
  GeV}},}\ }\href {\doibase 10.1103/PhysRevC.88.024906} {\bibfield  {journal}
  {\bibinfo  {journal} {Phys. Rev. C}\ }\textbf {\bibinfo {volume} {88}},\
  \bibinfo {pages} {024906} (\bibinfo {year} {2013}{\natexlab{d}})}\BibitemShut
  {NoStop}%
\bibitem [{\citenamefont {Adare}\ \emph
  {et~al.}(2013{\natexlab{e}})\citenamefont {Adare} \emph
  {et~al.}}]{Adare:2013ezl}%
  \BibitemOpen
  \bibfield  {author} {\bibinfo {author} {\bibfnamefont {A.}~\bibnamefont
  {Adare}} \emph {et~al.} (\bibinfo {collaboration} {PHENIX Collaboration}),\
  }\bibfield  {title} {\enquote {\bibinfo {title} {{Nuclear Modification of
  $\psi^{\prime}$, $\chi_c$, and $J/\psi$ Production in $d$+Au Collisions at
  $\sqrt{s_{NN}}$=200~GeV}},}\ }\href {\doibase 10.1103/PhysRevLett.111.202301}
  {\bibfield  {journal} {\bibinfo  {journal} {Phys. Rev. Lett.}\ }\textbf
  {\bibinfo {volume} {111}},\ \bibinfo {pages} {202301} (\bibinfo {year}
  {2013}{\natexlab{e}})}\BibitemShut {NoStop}%
\bibitem [{\citenamefont {Adare}\ \emph
  {et~al.}(2014{\natexlab{a}})\citenamefont {Adare} \emph
  {et~al.}}]{Adare:2013lkk}%
  \BibitemOpen
  \bibfield  {author} {\bibinfo {author} {\bibfnamefont {A.}~\bibnamefont
  {Adare}} \emph {et~al.} (\bibinfo {collaboration} {PHENIX Collaboration}),\
  }\bibfield  {title} {\enquote {\bibinfo {title} {{Cold-Nuclear-Matter Effects
  on Heavy-Quark Production at Forward and Backward Rapidity in $d$+Au
  Collisions at $\sqrt{s_{NN}}$=200~GeV}},}\ }\href {\doibase
  10.1103/PhysRevLett.112.252301} {\bibfield  {journal} {\bibinfo  {journal}
  {Phys. Rev. Lett.}\ }\textbf {\bibinfo {volume} {112}},\ \bibinfo {pages}
  {252301} (\bibinfo {year} {2014}{\natexlab{a}})}\BibitemShut {NoStop}%
\bibitem [{\citenamefont {Koch}\ \emph {et~al.}(1986)\citenamefont {Koch},
  \citenamefont {Muller},\ and\ \citenamefont {Rafelski}}]{Koch:1986ud}%
  \BibitemOpen
  \bibfield  {author} {\bibinfo {author} {\bibfnamefont {P.}~\bibnamefont
  {Koch}}, \bibinfo {author} {\bibfnamefont {B.}~\bibnamefont {Muller}}, \ and\
  \bibinfo {author} {\bibfnamefont {J.}~\bibnamefont {Rafelski}},\ }\bibfield
  {title} {\enquote {\bibinfo {title} {{Strangeness in Relativistic Heavy Ion
  Collisions}},}\ }\href {\doibase 10.1016/0370-1573(86)90096-7} {\bibfield
  {journal} {\bibinfo  {journal} {Phys.Rept.}\ }\textbf {\bibinfo {volume}
  {142}},\ \bibinfo {pages} {167} (\bibinfo {year} {1986})}\BibitemShut
  {NoStop}%
\bibitem [{\citenamefont {Shor}(1985)}]{Shor:1984ui}%
  \BibitemOpen
  \bibfield  {author} {\bibinfo {author} {\bibfnamefont {A.}~\bibnamefont
  {Shor}},\ }\bibfield  {title} {\enquote {\bibinfo {title} {{Phi meson
  production as a probe of the quark gluon plasma}},}\ }\href {\doibase
  10.1103/PhysRevLett.54.1122} {\bibfield  {journal} {\bibinfo  {journal}
  {Phys. Rev. Lett.}\ }\textbf {\bibinfo {volume} {54}},\ \bibinfo {pages}
  {1122} (\bibinfo {year} {1985})}\BibitemShut {NoStop}%
\bibitem [{\citenamefont {Alessandro}\ \emph {et~al.}(2003)\citenamefont
  {Alessandro} \emph {et~al.}}]{Alessandro:2003gy}%
  \BibitemOpen
  \bibfield  {author} {\bibinfo {author} {\bibfnamefont {B.}~\bibnamefont
  {Alessandro}} \emph {et~al.} (\bibinfo {collaboration} {NA50
  Collaboration}),\ }\bibfield  {title} {\enquote {\bibinfo {title} {{Phi
  production in Pb-Pb collisions at 158~GeV/$c$ per nucleon incident
  momentum}},}\ }\href {\doibase 10.1016/S0370-2693(02)03267-7} {\bibfield
  {journal} {\bibinfo  {journal} {Phys. Lett. B}\ }\textbf {\bibinfo {volume}
  {555}},\ \bibinfo {pages} {147} (\bibinfo {year} {2003})}\BibitemShut
  {NoStop}%
\bibitem [{\citenamefont {Adler}\ \emph {et~al.}(2005)\citenamefont {Adler}
  \emph {et~al.}}]{Adler:2004hv}%
  \BibitemOpen
  \bibfield  {author} {\bibinfo {author} {\bibfnamefont {S.~S.}\ \bibnamefont
  {Adler}} \emph {et~al.} (\bibinfo {collaboration} {PHENIX Collaboration}),\
  }\bibfield  {title} {\enquote {\bibinfo {title} {{Production of phi mesons at
  mid-rapidity in $\sqrt{s_{NN}}$=200~GeV Au+Au collisions at RHIC}},}\ }\href
  {\doibase 10.1103/PhysRevC.72.014903} {\bibfield  {journal} {\bibinfo
  {journal} {Phys. Rev. C}\ }\textbf {\bibinfo {volume} {72}},\ \bibinfo
  {pages} {014903} (\bibinfo {year} {2005})}\BibitemShut {NoStop}%
\bibitem [{\citenamefont {Afanasiev}\ \emph {et~al.}(2007)\citenamefont
  {Afanasiev} \emph {et~al.}}]{Afanasiev:2007tv}%
  \BibitemOpen
  \bibfield  {author} {\bibinfo {author} {\bibfnamefont {S.}~\bibnamefont
  {Afanasiev}} \emph {et~al.} (\bibinfo {collaboration} {PHENIX
  Collaboration}),\ }\bibfield  {title} {\enquote {\bibinfo {title} {{Elliptic
  flow for phi mesons and (anti)deuterons in Au + Au collisions at
  $\sqrt{s_{NN}}$=200~GeV}},}\ }\href {\doibase 10.1103/PhysRevLett.99.052301}
  {\bibfield  {journal} {\bibinfo  {journal} {Phys. Rev. Lett.}\ }\textbf
  {\bibinfo {volume} {99}},\ \bibinfo {pages} {052301} (\bibinfo {year}
  {2007})}\BibitemShut {NoStop}%
\bibitem [{\citenamefont {Abelev}\ \emph {et~al.}(2007)\citenamefont {Abelev}
  \emph {et~al.}}]{Abelev:2007rw}%
  \BibitemOpen
  \bibfield  {author} {\bibinfo {author} {\bibfnamefont {B.~I.}\ \bibnamefont
  {Abelev}} \emph {et~al.} (\bibinfo {collaboration} {STAR Collaboration}),\
  }\bibfield  {title} {\enquote {\bibinfo {title} {{Partonic flow and phi-meson
  production in Au+Au collisions at $\sqrt{s_{NN}}$ = 200~GeV}},}\ }\href
  {\doibase 10.1103/PhysRevLett.99.112301} {\bibfield  {journal} {\bibinfo
  {journal} {Phys. Rev. Lett.}\ }\textbf {\bibinfo {volume} {99}},\ \bibinfo
  {pages} {112301} (\bibinfo {year} {2007})}\BibitemShut {NoStop}%
\bibitem [{\citenamefont {Adamova}\ \emph {et~al.}(2008)\citenamefont {Adamova}
  \emph {et~al.}}]{Adamova:2006nu}%
  \BibitemOpen
  \bibfield  {author} {\bibinfo {author} {\bibfnamefont {D.}~\bibnamefont
  {Adamova}} \emph {et~al.} (\bibinfo {collaboration} {CERES Collaboration}),\
  }\bibfield  {title} {\enquote {\bibinfo {title} {{Modification of the
  rho-meson detected by low-mass electron-positron pairs in central Pb-Au
  collisions at 158-A~GeV/c}},}\ }\href {\doibase
  10.1016/j.physletb.2008.07.104} {\bibfield  {journal} {\bibinfo  {journal}
  {Phys. Lett. B}\ }\textbf {\bibinfo {volume} {666}},\ \bibinfo {pages} {425}
  (\bibinfo {year} {2008})}\BibitemShut {NoStop}%
\bibitem [{\citenamefont {Alt}\ \emph {et~al.}(2008)\citenamefont {Alt} \emph
  {et~al.}}]{Alt:2008iv}%
  \BibitemOpen
  \bibfield  {author} {\bibinfo {author} {\bibfnamefont {C.}~\bibnamefont
  {Alt}} \emph {et~al.} (\bibinfo {collaboration} {NA49 Collaboration}),\
  }\bibfield  {title} {\enquote {\bibinfo {title} {{Energy dependence of phi
  meson production in central Pb+Pb collisions at $\sqrt{s_{NN}}$=6 to 17
  GeV}},}\ }\href {\doibase 10.1103/PhysRevC.78.044907} {\bibfield  {journal}
  {\bibinfo  {journal} {Phys. Rev. C}\ }\textbf {\bibinfo {volume} {78}},\
  \bibinfo {pages} {044907} (\bibinfo {year} {2008})}\BibitemShut {NoStop}%
\bibitem [{\citenamefont {Abelev}\ \emph
  {et~al.}(2009{\natexlab{b}})\citenamefont {Abelev} \emph
  {et~al.}}]{Abelev:2008zk}%
  \BibitemOpen
  \bibfield  {author} {\bibinfo {author} {\bibfnamefont {B.~I.}\ \bibnamefont
  {Abelev}} \emph {et~al.} (\bibinfo {collaboration} {STAR Collaboration}),\
  }\bibfield  {title} {\enquote {\bibinfo {title} {{Energy and system size
  dependence of phi meson production in Cu+Cu and Au+Au collisions}},}\ }\href
  {\doibase 10.1016/j.physletb.2009.02.037} {\bibfield  {journal} {\bibinfo
  {journal} {Phys. Lett. B}\ }\textbf {\bibinfo {volume} {673}},\ \bibinfo
  {pages} {183} (\bibinfo {year} {2009}{\natexlab{b}})}\BibitemShut {NoStop}%
\bibitem [{\citenamefont {Adare}\ \emph
  {et~al.}(2011{\natexlab{a}})\citenamefont {Adare} \emph
  {et~al.}}]{Adare:2010pt}%
  \BibitemOpen
  \bibfield  {author} {\bibinfo {author} {\bibfnamefont {A.}~\bibnamefont
  {Adare}} \emph {et~al.} (\bibinfo {collaboration} {PHENIX Collaboration}),\
  }\bibfield  {title} {\enquote {\bibinfo {title} {{Nuclear modification
  factors of $\phi$ mesons in $d+$Au, Cu+Cu and Au+Au collisions at
  $\sqrt{s_{NN}}$=200 GeV}},}\ }\href {\doibase 10.1103/PhysRevC.83.024909}
  {\bibfield  {journal} {\bibinfo  {journal} {Phys. Rev. C}\ }\textbf {\bibinfo
  {volume} {83}},\ \bibinfo {pages} {024909} (\bibinfo {year}
  {2011}{\natexlab{a}})}\BibitemShut {NoStop}%
\bibitem [{\citenamefont {Arnaldi}\ \emph {et~al.}(2011)\citenamefont {Arnaldi}
  \emph {et~al.}}]{Arnaldi:2011nn}%
  \BibitemOpen
  \bibfield  {author} {\bibinfo {author} {\bibfnamefont {R.}~\bibnamefont
  {Arnaldi}} \emph {et~al.} (\bibinfo {collaboration} {NA60 Collaboration}),\
  }\bibfield  {title} {\enquote {\bibinfo {title} {{A Comparative measurement
  of $\phi\rightarrow K^+K^-$ and $\phi\rightarrow \mu^+\mu^-$ in In-In
  collisions at the CERN SPS}},}\ }\href {\doibase
  10.1016/j.physletb.2011.04.028} {\bibfield  {journal} {\bibinfo  {journal}
  {Phys. Lett. B}\ }\textbf {\bibinfo {volume} {699}},\ \bibinfo {pages} {325}
  (\bibinfo {year} {2011})}\BibitemShut {NoStop}%
\bibitem [{\citenamefont {Adare}\ \emph
  {et~al.}(2011{\natexlab{b}})\citenamefont {Adare} \emph
  {et~al.}}]{Adare:2010fe}%
  \BibitemOpen
  \bibfield  {author} {\bibinfo {author} {\bibfnamefont {A.}~\bibnamefont
  {Adare}} \emph {et~al.} (\bibinfo {collaboration} {PHENIX Collaboration}),\
  }\bibfield  {title} {\enquote {\bibinfo {title} {{Measurement of neutral
  mesons in $p+p$ collisions at $\sqrt{s}$= 200 GeV and scaling properties of
  hadron production}},}\ }\href {\doibase 10.1103/PhysRevD.83.052004}
  {\bibfield  {journal} {\bibinfo  {journal} {Phys. Rev. D}\ }\textbf {\bibinfo
  {volume} {83}},\ \bibinfo {pages} {052004} (\bibinfo {year}
  {2011}{\natexlab{b}})}\BibitemShut {NoStop}%
\bibitem [{\citenamefont {Adare}\ \emph
  {et~al.}(2014{\natexlab{b}})\citenamefont {Adare} \emph
  {et~al.}}]{Adare:2014mgt}%
  \BibitemOpen
  \bibfield  {author} {\bibinfo {author} {\bibfnamefont {A.}~\bibnamefont
  {Adare}} \emph {et~al.} (\bibinfo {collaboration} {PHENIX Collaboration}),\
  }\bibfield  {title} {\enquote {\bibinfo {title} {{Low-mass vector-meson
  production at forward rapidity in $p$$+$$p$ collisions at $\sqrt{s}=200$
  GeV}},}\ }\href {\doibase 10.1103/PhysRevD.90.052002} {\bibfield  {journal}
  {\bibinfo  {journal} {Phys. Rev. D}\ }\textbf {\bibinfo {volume} {90}},\
  \bibinfo {pages} {052002} (\bibinfo {year} {2014}{\natexlab{b}})}\BibitemShut
  {NoStop}%
\bibitem [{\citenamefont {Adcox}\ \emph {et~al.}(2003)\citenamefont {Adcox}
  \emph {et~al.}}]{Adcox:2003zm}%
  \BibitemOpen
  \bibfield  {author} {\bibinfo {author} {\bibfnamefont {K.}~\bibnamefont
  {Adcox}} \emph {et~al.} (\bibinfo {collaboration} {PHENIX Collaboration}),\
  }\bibfield  {title} {\enquote {\bibinfo {title} {{PHENIX detector
  overview}},}\ }\href {\doibase 10.1016/S0168-9002(02)01950-2} {\bibfield
  {journal} {\bibinfo  {journal} {Nucl. Instrum. Methods Phys. Res., Sec. A}\
  }\textbf {\bibinfo {volume} {499}},\ \bibinfo {pages} {469} (\bibinfo {year}
  {2003})}\BibitemShut {NoStop}%
\bibitem [{\citenamefont {Beringer}\ \emph {et~al.}(2012)\citenamefont
  {Beringer} \emph {et~al.}}]{Beringer:1900zz}%
  \BibitemOpen
  \bibfield  {author} {\bibinfo {author} {\bibfnamefont {J.}~\bibnamefont
  {Beringer}} \emph {et~al.} (\bibinfo {collaboration} {Particle Data Group}),\
  }\bibfield  {title} {\enquote {\bibinfo {title} {{Rev. of Particle Phys.
  (RPP)}},}\ }\href {\doibase 10.1103/PhysRevD.86.010001} {\bibfield  {journal}
  {\bibinfo  {journal} {Phys. Rev. D}\ }\textbf {\bibinfo {volume} {86}},\
  \bibinfo {pages} {010001} (\bibinfo {year} {2012})}\BibitemShut {NoStop}%
\bibitem [{\citenamefont {Yao}\ \emph {et~al.}(2006)\citenamefont {Yao} \emph
  {et~al.}}]{JPhysG33:pdg1}%
  \BibitemOpen
  \bibfield  {author} {\bibinfo {author} {\bibfnamefont {W.~M.}\ \bibnamefont
  {Yao}} \emph {et~al.} (\bibinfo {collaboration} {Particle Data Group}),\
  }\bibfield  {title} {\enquote {\bibinfo {title} {{Review of Particle
  Physics}},}\ }\href@noop {} {\bibfield  {journal} {\bibinfo  {journal} {J.
  Phys. G}\ }\textbf {\bibinfo {volume} {33}},\ \bibinfo {pages} {1} (\bibinfo
  {year} {2006})}\BibitemShut {NoStop}%
\bibitem [{\citenamefont {Sjostrand}\ \emph {et~al.}(2001)\citenamefont
  {Sjostrand}, \citenamefont {Eden}, \citenamefont {Friberg}, \citenamefont
  {Lonnblad}, \citenamefont {Miu}, \citenamefont {Mrenna},\ and\ \citenamefont
  {Norrbin}}]{Sjostrand:2000wi}%
  \BibitemOpen
  \bibfield  {author} {\bibinfo {author} {\bibfnamefont {T.}~\bibnamefont
  {Sjostrand}}, \bibinfo {author} {\bibfnamefont {P.}~\bibnamefont {Eden}},
  \bibinfo {author} {\bibfnamefont {C.}~\bibnamefont {Friberg}}, \bibinfo
  {author} {\bibfnamefont {L.}~\bibnamefont {Lonnblad}}, \bibinfo {author}
  {\bibfnamefont {G.}~\bibnamefont {Miu}}, \bibinfo {author} {\bibfnamefont
  {S.}~\bibnamefont {Mrenna}}, \ and\ \bibinfo {author} {\bibfnamefont
  {E.}~\bibnamefont {Norrbin}},\ }\bibfield  {title} {\enquote {\bibinfo
  {title} {{High-energy physics event generation with {\sc pythia} 6.1}},}\
  }\href {\doibase 10.1016/S0010-4655(00)00236-8} {\bibfield  {journal}
  {\bibinfo  {journal} {Comput. Phys. Commun.}\ }\textbf {\bibinfo {volume}
  {135}},\ \bibinfo {pages} {238} (\bibinfo {year} {2001})}\BibitemShut
  {NoStop}%
\bibitem [{GEA(1994)}]{GEANT:W5013}%
  \BibitemOpen
  \href {http://wwwasdoc.web.cern.ch/wwwasdoc/pdfdir/geant.pdf} {\emph
  {\bibinfo {title} {GEANT 3.2.1 Manual}}} (\bibinfo {year} {1994}),\ \bibinfo
  {note} {{CERN W5013}}\BibitemShut {NoStop}%
\bibitem [{\citenamefont {Adare}\ \emph
  {et~al.}(2011{\natexlab{c}})\citenamefont {Adare} \emph
  {et~al.}}]{Adare:2011ht}%
  \BibitemOpen
  \bibfield  {author} {\bibinfo {author} {\bibfnamefont {A.}~\bibnamefont
  {Adare}} \emph {et~al.} (\bibinfo {collaboration} {PHENIX Collaboration}),\
  }\bibfield  {title} {\enquote {\bibinfo {title} {{Production of $\omega$
  mesons in $p$+$p$, $d$+Au, Cu+Cu, and Au+Au collisions at $\sqrt{s_{NN}}=200$
  GeV}},}\ }\href {\doibase 10.1103/PhysRevC.84.044902} {\bibfield  {journal}
  {\bibinfo  {journal} {Phys. Rev. C}\ }\textbf {\bibinfo {volume} {84}},\
  \bibinfo {pages} {044902} (\bibinfo {year} {2011}{\natexlab{c}})}\BibitemShut
  {NoStop}%
\bibitem [{\citenamefont {Adare}\ \emph {et~al.}(2010)\citenamefont {Adare}
  \emph {et~al.}}]{Adare:2009qk}%
  \BibitemOpen
  \bibfield  {author} {\bibinfo {author} {\bibfnamefont {A.}~\bibnamefont
  {Adare}} \emph {et~al.} (\bibinfo {collaboration} {PHENIX Collaboration}),\
  }\bibfield  {title} {\enquote {\bibinfo {title} {{Detailed measurement of the
  $e^+ e^-$ pair continuum in $p+p$ and Au+Au collisions at $\sqrt{s_{NN}}$=200
  GeV and implications for direct photon production}},}\ }\href {\doibase
  10.1103/PhysRevC.81.034911} {\bibfield  {journal} {\bibinfo  {journal} {Phys.
  Rev. C}\ }\textbf {\bibinfo {volume} {81}},\ \bibinfo {pages} {034911}
  (\bibinfo {year} {2010})}\BibitemShut {NoStop}%
\bibitem [{\citenamefont {Adare}\ \emph
  {et~al.}(2014{\natexlab{c}})\citenamefont {Adare} \emph
  {et~al.}}]{Adare:2013nff}%
  \BibitemOpen
  \bibfield  {author} {\bibinfo {author} {\bibfnamefont {A.}~\bibnamefont
  {Adare}} \emph {et~al.} (\bibinfo {collaboration} {PHENIX Collaboration}),\
  }\bibfield  {title} {\enquote {\bibinfo {title} {{Centrality categorization
  for $R_{p(d)+A}$ in high-energy collisions}},}\ }\href {\doibase
  10.1103/PhysRevC.90.034902} {\bibfield  {journal} {\bibinfo  {journal} {Phys.
  Rev. C}\ }\textbf {\bibinfo {volume} {90}},\ \bibinfo {pages} {034902}
  (\bibinfo {year} {2014}{\natexlab{c}})}\BibitemShut {NoStop}%
\bibitem [{\citenamefont {Adler}\ \emph {et~al.}(2006)\citenamefont {Adler}
  \emph {et~al.}}]{Adler:2005ph}%
  \BibitemOpen
  \bibfield  {author} {\bibinfo {author} {\bibfnamefont {S.~S.}\ \bibnamefont
  {Adler}} \emph {et~al.} (\bibinfo {collaboration} {PHENIX Collaboration}),\
  }\bibfield  {title} {\enquote {\bibinfo {title} {{$J/\psi$ production and
  nuclear effects for $d$$+$Au and $p$$+$$p$ collisions at
  $\sqrt{s_{NN}}$=200~GeV}},}\ }\href {\doibase 10.1103/PhysRevLett.96.012304}
  {\bibfield  {journal} {\bibinfo  {journal} {Phys. Rev. Lett.}\ }\textbf
  {\bibinfo {volume} {96}},\ \bibinfo {pages} {012304} (\bibinfo {year}
  {2006})}\BibitemShut {NoStop}%
\bibitem [{\citenamefont {Lafferty}\ and\ \citenamefont
  {Wyatt}(1995)}]{Lafferty:1994cj}%
  \BibitemOpen
  \bibfield  {author} {\bibinfo {author} {\bibfnamefont {G.~D.}\ \bibnamefont
  {Lafferty}}\ and\ \bibinfo {author} {\bibfnamefont {T.R.}\ \bibnamefont
  {Wyatt}},\ }\bibfield  {title} {\enquote {\bibinfo {title} {{Where to stick
  your data points: The treatment of measurements within wide bins}},}\ }\href
  {\doibase 10.1016/0168-9002(94)01112-5} {\bibfield  {journal} {\bibinfo
  {journal} {Nucl. Instrum. Methods Phys. Res., Sec. A}\ }\textbf {\bibinfo
  {volume} {355}},\ \bibinfo {pages} {541} (\bibinfo {year}
  {1995})}\BibitemShut {NoStop}%
\bibitem [{\citenamefont {Hagedorn}(1965)}]{Hagedorn:1965st}%
  \BibitemOpen
  \bibfield  {author} {\bibinfo {author} {\bibfnamefont {R.}~\bibnamefont
  {Hagedorn}},\ }\bibfield  {title} {\enquote {\bibinfo {title} {{Statistical
  thermodynamics of strong interactions at high-energies}},}\ }\href@noop {}
  {\bibfield  {journal} {\bibinfo  {journal} {Nuovo Cim. Suppl.}\ }\textbf
  {\bibinfo {volume} {3}},\ \bibinfo {pages} {147} (\bibinfo {year}
  {1965})}\BibitemShut {NoStop}%
\bibitem [{\citenamefont {Kaplan}\ \emph {et~al.}(1978)\citenamefont {Kaplan}
  \emph {et~al.}}]{Kaplan:1977kr}%
  \BibitemOpen
  \bibfield  {author} {\bibinfo {author} {\bibfnamefont {D.~M.}\ \bibnamefont
  {Kaplan}} \emph {et~al.} (\bibinfo {collaboration} {FNAL-E288
  Collaboration}),\ }\bibfield  {title} {\enquote {\bibinfo {title} {{Study of
  the High Mass Dimuon Continuum in 400~GeV Proton-Nucleus Collisions}},}\
  }\href {\doibase 10.1103/PhysRevLett.40.435} {\bibfield  {journal} {\bibinfo
  {journal} {Phys. Rev. Lett.}\ }\textbf {\bibinfo {volume} {40}},\ \bibinfo
  {pages} {435} (\bibinfo {year} {1978})}\BibitemShut {NoStop}%
\bibitem [{\citenamefont {Adare}\ \emph
  {et~al.}(2011{\natexlab{d}})\citenamefont {Adare} \emph
  {et~al.}}]{Adare:2010fn}%
  \BibitemOpen
  \bibfield  {author} {\bibinfo {author} {\bibfnamefont {A.}~\bibnamefont
  {Adare}} \emph {et~al.} (\bibinfo {collaboration} {PHENIX Collaboration}),\
  }\bibfield  {title} {\enquote {\bibinfo {title} {{Cold Nuclear Matter Effects
  on $J/\psi$ Yields as a Function of Rapidity and Nuclear Geometry in
  Deuteron-Gold Collisions at $\sqrt{s_{NN}}$=200 GeV}},}\ }\href {\doibase
  10.1103/PhysRevLett.107.142301} {\bibfield  {journal} {\bibinfo  {journal}
  {Phys. Rev. Lett.}\ }\textbf {\bibinfo {volume} {107}},\ \bibinfo {pages}
  {142301} (\bibinfo {year} {2011}{\natexlab{d}})}\BibitemShut {NoStop}%
\bibitem [{\citenamefont {Adam}\ \emph {et~al.}()\citenamefont {Adam} \emph
  {et~al.}}]{Adam:2015jca}%
  \BibitemOpen
  \bibfield  {author} {\bibinfo {author} {\bibfnamefont {J.}~\bibnamefont
  {Adam}} \emph {et~al.} (\bibinfo {collaboration} {ALICE Collaboration}),\
  }\href@noop {} {\enquote {\bibinfo {title} {{phi-meson production at forward
  rapidity in $p+Pb$ collisions at $\sqrt{s_{NN}}$=5.02 TeV and in $p+p$
  collisions at $\sqrt{s}$=2.86 TeV}},}\ }\bibinfo {note}
  {{arXiv:1506.09206}}\BibitemShut {NoStop}%
\end{thebibliography}

%
 
\end{document}